\documentstyle[nato,epsfig,rotate,axodraw,graphicx,pstricks,pst-coil,
psfig]{crckapb}

\def\a{\alpha}
\def\b{\beta}

\def\d{\delta}
\def\e{\epsilon}

\def\g{\gamma}

\def\j{\psi}
\def\k{\kappa}
\def\l{\lambda}
\def\m{\mu}
\def\n{\nu}

\def\p{\pi}

\def\r{\rho}
\def\s{\sigma}
\def\t{\tau}

\def\z{\zeta}
\def\D{\Delta}
\def\F{\Phi}
\def\G{\Gamma}

\def\L{\Lambda}
\def\O{\Omega}

\def\Q{\Theta}

\def\ve{\varepsilon}
\def\vf{\varphi}

\def\co{{\cal O}}
\def\cm{{\cal M}}
\def\cl{{\cal L}}

\def\wt#1{\widetilde{#1}}                    
\def\dg{\dagger}                                     
\def\VEV#1{\left\langle #1\right\rangle}        
\def\Bar#1{\overline{#1}}                       
\def\beq{\begin{equation}}
\def\eeq{\end{equation}}
\def\bea{\begin{eqnarray}}
\def\eea{\end{eqnarray}}
\def\NO{\nonumber}
\newcommand{\gr}{\tilde{G}}
\newcommand{\gl}{\tilde{g}}                     
\newcommand{\factor}{\left(1 + \frac{\mgl^2}{3\mgr^2} \right)}
\newcommand{\mgr}{m_{\gr}}
\newcommand{\mgl}{m_{\gl}}
\newcommand{\ngr}{n_{\gr}}
\newcommand{\nrad}{n_{\mathrm{rad}}}
\newcommand{\ttt}{2\rightarrow 2}
\newcommand{\Ygr}{Y_{\gr}}
\newcommand{\Ogr}{\O_{\gr}}

\def\pl#1#2#3{Phys.~Lett.~{\bf B {#1}} (19{#2}) #3}
\def\np#1#2#3{Nucl.~Phys.~{\bf B {#1}} (19{#2}) #3}
\def\prl#1#2#3{Phys.~Rev.~Lett.~{\bf #1} (19{#2}) #3}
\def\pr#1#2#3{Phys.~Rev.~{\bf D {#1}} (19{#2}) #3}

\def\ptp#1#2#3{Progr.~Theor.~Phys.~{\bf {#1}} (19{#2}) #3}

\def\nc#1#2#3{Nuovo Cim.~{\bf {#1}} (19{#2}) #3}

\begin{opening}

\title{SOME ASPECTS OF BARYOGENESIS AND }

\subtitle{\rm\bf LEPTON NUMBER VIOLATION }

\author{W. BUCHMULLER}

\institute{Deutsches Elektronen-Synchrotron DESY\\
Notkestr.\ 85, D--22607 Hamburg, Germany}

\end{opening}

\runningtitle{ASPECTS OF BARYOGENESIS}

\begin{document}


\begin{abstract}
The cosmological baryon asymmetry is closely related to neutrino properties
due to the non-perturbative sphaleron processes in the high-temperature
symmetric phase of the standard model. We review some aspects of this
connection with emphasis on cosmological bounds on neutrino masses,
leptogenesis and possible implications for dark matter.
\end{abstract}

\vspace*{-11.5truecm}
\begin{flushright}
\begin{tabular}{l}
DESY 00--194\\
December 2000
\end{tabular}
\end{flushright}
\vspace*{9.5truecm}
\begin{center}
{\small{\it Lectures at the NATO ASI 2000, Cascais, Portugal, 
26 June -- 7 July, 2000}}
\end{center}
\vspace*{-0.2truecm}

\section{Elements of baryogenesis}

Particle physics unravels the structure of matter at short distances. The
characteristic energy scales of the different layers of structure are of order 
1~eV, the binding energy of atoms, 1~MeV, the binding energy of nuclei, and 
100~GeV, the typical collision energy of quarks 
and leptons in present day high-energy colliders. Knowledge of the laws of
nature which govern the interactions of elementary particles at these energies
allows us to calculate the properties of a plasma of particles at the 
corresponding temperatures. Extrapolating the observed Hubble expansion of the
universe back to early times one concludes that such temperatures must have
been realized in the early stages of the evolution of the universe.

At a temperature $T \sim 1$~eV electrons and nuclei combined to neutral atoms
and the universe became transparent to photons. The observation and recent
detailed investigation of the corresponding cosmic microwave background
\cite{boo} is the basis of early-universe cosmology. The second main
success is the prediction of the abundances of the light elements, D, $^3$He, 
$^4$He and $^7$Li based on the properties of the lightest baryons, protons 
and neutrons. The light 
elements were formed at a temperature $T \sim 1$~MeV. Agreement between theory
and observation is obtained for a certain range of the parameter $\eta$, the 
ratio of baryon density and photon density \cite{rpp00},
\beq
\eta = {n_B\over n_\g} = (1.5 - 6.3)\times 10^{-10}\;,
\eeq
where the present number density of photons is $n_\g \sim 400/{\rm cm}^3$. 
Since no significant amount of antimatter is observed in the universe, 
the baryon density is equal to the cosmological baryon asymmetry, 
$n_B \simeq n_B - n_{\bar{B}}$.

The formation of the cosmological baryon asymmetry, i.e., the origin of
the observed matter density today, can be understood in terms of the 
properties of quarks and leptons, the building blocks of the standard model
which describes their interactons at energies of order 100 GeV
and beyond. There are three families of left-handed and right-handed up-type
and down-type quarks with baryon number $B=1/3$ and lepton number $L=0$,
and three families of left-handed neutrinos and left- and right-handed
electron-type leptons with $B=0$ and $L=1$,
\beq
q_{Li}=\left(\begin{array}{c} u_{Li} \\ d_{Li} \end{array}\right)\ , 
\; u_{Ri}\ , \; d_{Ri}\ ,\quad                  
l_{Li}=\left(\begin{array}{c} \n_{Li} \\ e_{Li} \end{array}\right)\ , 
\; e_{Ri}\ ,\quad i=1\ldots 3\;.
\eeq
Some of their interactions are not invariant under charge conjugation (C) and
the combined charge conjugation and parity (CP) transformation \cite{RF}.

A matter-antimatter asymmetry can be dynamically generated in an expanding
universe if the particle interactions and the cosmological evolution satisfy 
Sakharov's conditions \cite{sa67}, i.e.,
\begin{itemize}
\item baryon number violation
\item $C$ and $C\!P$ violation
\item deviation from thermal equilibrium .
\end{itemize}
At present there are a number of viable scenarios for baryogenesis. They
can be classified according to the different ways in which Sakharov's 
conditions are realized. Already in the standard model $C$ and $C\!P$ are
not conserved. Also $B$ and $L$ are
violated by instanton processes \cite{tho76}. In grand unified theories
$B$ and $L$ are broken by the interactions of heavy gauge bosons and leptoquarks.
This is the basis of the classical GUT baryogenesis \cite{yo78}.
Analogously, the $L$ violating decays of heavy Majorana neutrinos lead to
leptogenesis \cite{fy86}. In supersymmetric theories the existence of 
approximately flat directions in the scalar potential provides new 
possibilities. Coherent oscillations of scalar fields may then generate
large asymmetries \cite{ad85}.

The crucial departure from thermal equilibrium can also be realized in several
ways. One possibility is a sufficiently strong first-order electroweak phase 
transition \cite{CW}. In this case $C\!P$ violating interactions of the
standard model or its supersymmetric extension could in principle generate the
observed baryon asymmetry. However, due to the rather large lower bound on the
Higgs boson mass of about 115~GeV, which is imposed by the LEP experiments,
this interesting possibility is now restricted to a very small range of 
parameters in the supersymmetric standard model. In the case of the 
Affleck-Dine scenario the baryon asymmetry is generated at the end of an 
inflationary period as a coherent effect of scalar fields which leads to an
asymmetry between quarks and antiquarks after reheating \cite{AL}.
For the classical GUT baryogenesis and for leptogenesis the departure from
thermal equilibrium is due to the deviation of the number density of the
decaying heavy particles from the equilibrium number density.
How strong this deviation from thermal equilibrium is depends on the lifetime
of the decaying heavy particles and the cosmological evolution. Further
scenarios for baryogenesis are described in \cite{rt99}.

The theory of baryogenesis involves non-perturbative aspects of quantum
field theory and also non-equilibrium statistical field theory, in particular
the theory of phase transitions and kinetic theory. 
\begin{figure}[t]
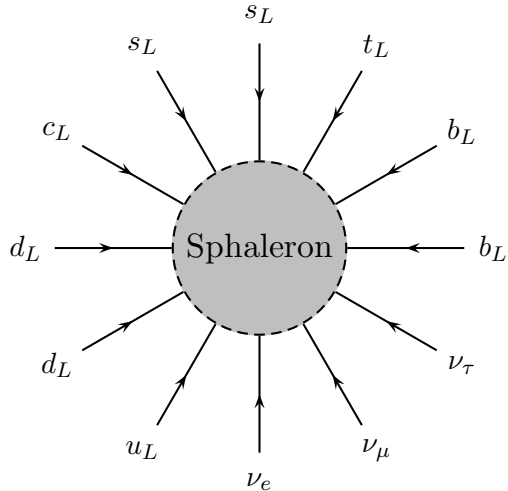

\begin{center}
\scaleboxto(7,7) {\parbox[c]{9cm}{ \begin{center}
     \pspicture*(-0.50,-2.5)(8.5,6.5)
     \psset{linecolor=lightgray}
     \qdisk(4,2){1.5cm}
     \psset{linecolor=black}
     \pscircle[linewidth=1pt,linestyle=dashed](4,2){1.5cm}
     \rput[cc]{0}(4,2){\scalebox{1.5}{Sphaleron}}
     \psline[linewidth=1pt](5.50,2.00)(7.50,2.00)
     \psline[linewidth=1pt](5.30,2.75)(7.03,3.75)
     \psline[linewidth=1pt](4.75,3.30)(5.75,5.03)
     \psline[linewidth=1pt](4.00,3.50)(4.00,5.50)
     \psline[linewidth=1pt](3.25,3.30)(2.25,5.03)
     \psline[linewidth=1pt](2.70,2.75)(0.97,3.75)
     \psline[linewidth=1pt](2.50,2.00)(0.50,2.00)
     \psline[linewidth=1pt](2.70,1.25)(0.97,0.25)
     \psline[linewidth=1pt](3.25,0.70)(2.25,-1.03)
     \psline[linewidth=1pt](4.00,0.50)(4.00,-1.50)
     \psline[linewidth=1pt](4.75,0.70)(5.75,-1.03)
     \psline[linewidth=1pt](5.30,1.25)(7.03,0.25)
     \psline[linewidth=1pt]{<-}(6.50,2.00)(6.60,2.00)
     \psline[linewidth=1pt]{<-}(6.17,3.25)(6.25,3.30)
     \psline[linewidth=1pt]{<-}(5.25,4.17)(5.30,4.25)
     \psline[linewidth=1pt]{<-}(4.00,4.50)(4.00,4.60)
     \psline[linewidth=1pt]{<-}(2.75,4.17)(2.70,4.25)
     \psline[linewidth=1pt]{<-}(1.83,3.25)(1.75,3.30)
     \psline[linewidth=1pt]{<-}(1.50,2.00)(1.40,2.00)
     \psline[linewidth=1pt]{<-}(1.83,0.75)(1.75,0.70)
     \psline[linewidth=1pt]{<-}(2.75,-0.17)(2.70,-0.25)
     \psline[linewidth=1pt]{<-}(4.00,-0.50)(4.00,-0.60)
     \psline[linewidth=1pt]{<-}(5.25,-0.17)(5.30,-0.25)
     \psline[linewidth=1pt]{<-}(6.17,0.75)(6.25,0.70)
     \rput[cc]{0}(8.00,2.00){\scalebox{1.3}{$b_L$}}
     \rput[cc]{0}(7.46,4.00){\scalebox{1.3}{$b_L$}}
     \rput[cc]{0}(6.00,5.46){\scalebox{1.3}{$t_L$}}
     \rput[cc]{0}(4.00,6.00){\scalebox{1.3}{$s_L$}}
     \rput[cc]{0}(2.00,5.46){\scalebox{1.3}{$s_L$}}
     \rput[cc]{0}(0.54,4.00){\scalebox{1.3}{$c_L$}}
     \rput[cc]{0}(0.00,2.00){\scalebox{1.3}{$d_L$}}
     \rput[cc]{0}(0.54,0.00){\scalebox{1.3}{$d_L$}}
     \rput[cc]{0}(2.00,-1.46){\scalebox{1.3}{$u_L$}}
     \rput[cc]{0}(4.00,-2.00){\scalebox{1.3}{$\nu_e$}}
     \rput[cc]{0}(6.00,-1.46){\scalebox{1.3}{$\nu_{\mu}$}}
     \rput[cc]{0}(7.46,0.00){\scalebox{1.3}{$\nu_{\tau}$}}
     \endpspicture
\end{center}}}
\end{center}
\caption{\it One of the 12-fermion processes which are in thermal 
equilibrium in the high-temperature phase of the standard model.
\label{fig_sphal}}
\end{figure}
A crucial ingredient is also the connection between baryon number and lepton 
number in the high-temperature, symmetric phase of
the standard model. Due to the chiral nature of the weak interactions $B$ and
$L$ are not conserved. At zero temperature this has no observable 
effect due to the smallness of the weak coupling. However, as the temperature 
approaches the critical temperature $T_{EW}$ of the electroweak transition, 
$B$ and $L$ violating processes come into thermal equilibrium \cite{krs85}. 

The rate of these processes is
related to the free energy of sphaleron-type field configurations which carry
topological charge. In the standard model they lead to an effective
interaction of all left-handed quarks and leptons \cite{tho76} 
(cf. fig.~\ref{fig_sphal}), 
\begin{equation}\label{obl}
O_{B+L} = \prod_i \left(q_{Li} q_{Li} q_{Li} l_{Li}\right)\; ,
\end{equation}
which violates baryon and lepton number by three units, 
\begin{equation} 
    \D B = \D L = 3\;. \label{sphal1}
\end{equation}

\begin{figure}[t]
\begin{center}
\epsfig{file=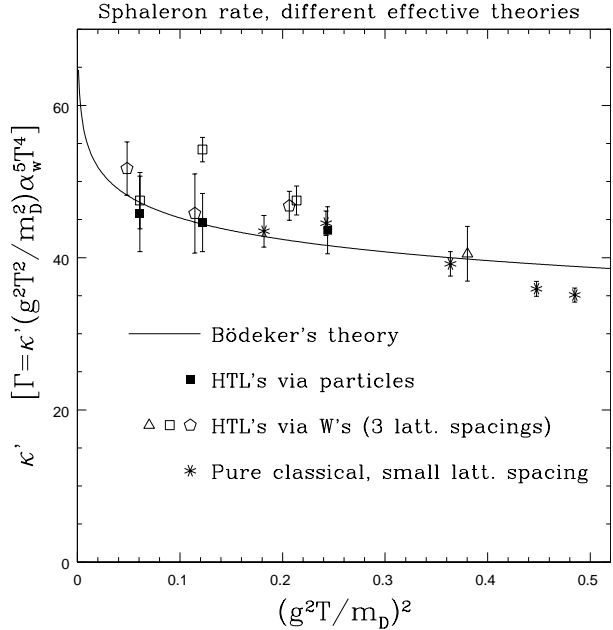,width=8cm,clip=}
\caption{{\it Sphaleron rate in B{\"o}deker's effective theory, two lattice 
implementations of HTL effective theory, and pure lattice theory interpreted
as HTL effective theory} \protect\cite{moo00}.}
\label{boed_fig}
\end{center}
\end{figure}
The evaluation of the sphaleron rate in the symmetric high-temperature phase 
is a complicated problem. A clear physical picture has been obtained in
B\"odeker's effective theory \cite{boe98} according to which low-frequency 
gauge field fluctuations satisfy the equation of motion
\beq
{\bf D}\times {\bf B} = \s {\bf E} - {\bf \z}\; .
\eeq
Here $\bf \z$ is Gaussian noise, i.e., a random vector field with variance
\beq
\langle \z_i(t,{\bf x})\z_j(t',{\bf x'})\rangle = 
2\s \d_{ij}\d(t-t')\d({\bf x}-{\bf x'})\;,
\eeq 
and $\s$ is a non-abelian conductivity. The sphaleron rate can then be 
written as \cite{moo00},
\beq
\G \simeq (14.0 \pm 0.3) {1\over \s} (\a_w T)^5\;.
\eeq
A comparison with two lattice simulations is shown in fig.~\ref{boed_fig}.
From this one derives that $B$ and $L$ violating processes are in thermal
equilibrium for temperatures in the range
\begin{equation}
T_{EW} \sim 100\ \mbox{GeV} < T < T_{SPH} \sim 10^{12}\ \mbox{GeV}\;.
\end{equation}

\begin{figure}
\begin{center}
\scaleboxto(7,3.5){
\parbox[c]{9cm}{\input{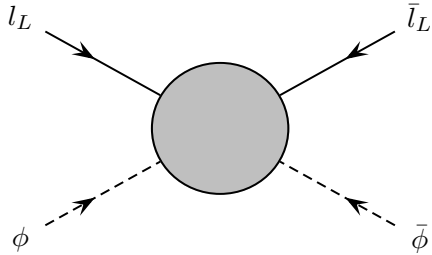}}}
\end{center}
\caption{\it Effective lepton number violating interaction.\label{fig_lept}}
\end{figure}

Sphaleron processes have a profound effect on the generation of the
cosmological baryon asymmetry, in particular in connection with lepton
number violating interactions between lepton and Higgs fields,
\begin{equation}\label{dl2}
\cl_{\Delta L=2} ={1\over 2} f_{ij}\ l^T_{Li}\vf\ C\ l_{Lj}\vf 
                  +\mbox{ h.c.}\;.\label{intl2}
\end{equation}
Such an interaction arises in particular from the exchange of heavy Majorana
neutrinos (cf.~fig.~\ref{fig_lept}). In the Higgs phase of the standard
model, where the Higgs field acquires a vacuum expectation value, it gives
rise to Majorana masses of the light neutrinos $\n_e$, $\n_\m$ and $\n_\t$.   

Eq.~(\ref{sphal1}) suggests that any
$B+L$ asymmetry generated before the electroweak phase transition,
i.e., at temperatures $T>T_{EW}$, will be washed out. However, since
only left-handed fields couple to sphalerons, a non-zero value of
$B+L$ can persist in the high-temperature, symmetric phase if there
exists a non-vanishing $B-L$ asymmetry. An analysis of the chemical potentials
of all particle species in the high-temperature phase yields a
relation between the baryon asymmetry $Y_B = (n_B-n_{\bar{B}})/s$, where $s$
is the entropy density, and the corresponding $B-L$ and $L$ asymmetries 
$Y_{B-L}$ and $Y_L$, respectively~\cite{ht90},
\beq\label{basic}
Y_B\ =\ a\ Y_{B-L}\ =\ {a\over a-1}\ Y_L\;.
\eeq
The number $a$ depends on the other processes which are in thermal 
equilibrium. If these are all standard model interactions one has 
$a=28/79$. If instead of the Yukawa interactions
of the right-handed electron the $\D L=2$ interactions (\ref{dl2}) are in
equilibrium one finds $a=-2/3$ ~\cite{bp00}.

The interplay between the sphaleron processes (fig.~\ref{fig_sphal}) 
and the lepton number violating processes (fig.~\ref{fig_lept}) leads to an
intriguing relation between neutrino properties and the cosmological baryon 
asymmetry. The decay of heavy Majorana neutrinos can quantitatively account
for the observed asymmetry.

\section{Heavy particle decays in a thermal bath}

Let us now consider the simplest possibility for a departure from thermal
equilibrium, the decay of heavy, weakly interacting particles in a thermal
bath. To be specific, we choose the heavy particle to be a
Majorana neutrino $N=N^c$ which can decay into a lepton Higgs  
pair $l \phi$ and also into the $C\!P$ conjugate state $\bar{l} \bar{\phi}$\,
\beq
N \rightarrow l\;\phi\;, \quad N \rightarrow \bar{l}\;\bar{\phi}\;.
\eeq
In the case of $C\!P$ violating couplings a lepton asymmetry can be generated in
the decays of the heavy neutrinos $N$ which is then partially transformed
into a baryon asymmetry by sphaleron processes \cite{fy86}. 
Compared to other scenarios of baryogenesis this leptogenesis mechanism
has the advantage that, at least in principle, the resulting baryon asymmetry
is entirely determined by neutrino properties. 

\begin{figure}[ht]
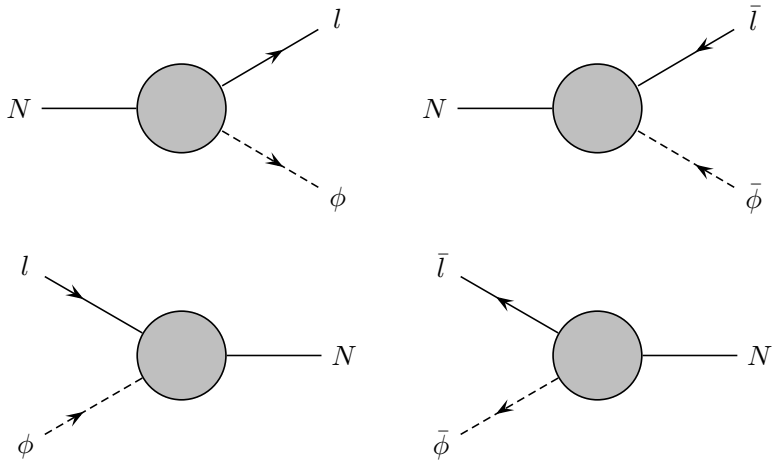

\begin{center}
\scaleboxto(5.4,0){\parbox[c]{9cm}{\input{decay1.tex}}}
\scaleboxto(5.4,0){\parbox[c]{9cm}{\input{decay2.tex}}}
\scaleboxto(5.4,0){\parbox[c]{9cm}{\input{decay3.tex}}}
\scaleboxto(5.4,0){\parbox[c]{9cm}{\input{decay4.tex}}}
\end{center}
\caption{\it $\D L=1$ processes: decays and inverse decays of a heavy Majorana
neutrino.\label{decinv}}
\end{figure}
\begin{figure}[ht]
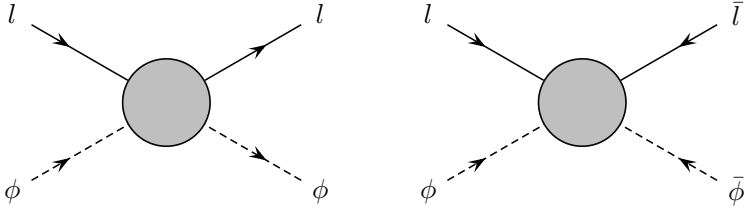

\begin{center}
\scaleboxto(5.4,0){
\parbox[c]{9cm}{\input{twotwo.tex}}}
\scaleboxto(5.4,0){
\parbox[c]{9cm}{\input{twoanti.tex}}}
\end{center}
\caption{\it $\D L=0$ and $\D L=2$ lepton Higgs processes.\label{lephig}}
\end{figure}

\subsection{Boltzmann equations}

The generation of a baryon asymmetry is an out-of-equilibrium process 
which is generally treated by means of Boltzmann equations. A detailed
discussion of the basic ideas and some of the subtleties has been given
in \cite{kw80}. The main processes in the thermal bath are the decays
and the inverse decays of the heavy neutrinos (cf.~fig.~\ref{decinv}), and
the lepton number conserving ($\D L=0$) and violating ($\D L=2$) processes
(cf.~fig.~\ref{lephig}). In addition there are other processes, in particular
those involving the t-quark, which are also important in a quantitative
analysis \cite{lut92,plu97}. A lepton asymmetry can be dynamically generated
in an expanding universe if the partial decay widths of the heavy neutrino
violate $C\!P$ invariance, i.e.,
\beq\label{gcp} 
\G(N\rightarrow l\phi) = {1\over 2}(1+\e)\G\;,\quad
\G(N\rightarrow \bar{l}\bar{\phi})={1\over 2}(1-\e)\G\;.
\eeq
Here $\G$ is the total decay width, and the parameter $\e \ll 1$ measures the 
amount of $C\!P$ violation.

The Boltzmann equations for the number densities of heavy neutrinos ($n_N$),
leptons ($n_l$) and antileptons ($n_{\bar{l}}$) corresponding to the 
processes in figs.~\ref{decinv} and \ref{lephig} are given by
\bea
{d n_N\over dt} + 3 H n_N &=& -\g(N \rightarrow l\phi) 
                              + \g(l\phi \rightarrow N) \NO\\
&&  -\g(N \rightarrow \bar{l}\bar{\phi})
    +\g(\bar{l}\bar{\phi} \rightarrow N)\;,\label{nN}\\
{d n_l\over dt} + 3 H n_l &=& \g(N \rightarrow l\phi) 
                              - \g(l\phi \rightarrow N) \NO\\
&& + \g(\bar{l}\bar{\phi}\rightarrow l\phi)
   - \g(l\phi \rightarrow \bar{l}\bar{\phi})\;,\label{nl}\\
{d n_{\bar{l}}\over dt} + 3 H n_{\bar{l}} 
&=& \g(N \rightarrow \bar{l}\bar{\phi})
    -\g(\bar{l}\bar{\phi}\rightarrow N) \NO\\
&&  +\g(l\phi \rightarrow \bar{l}\bar{\phi})
    - \g(\bar{l}\bar{\phi}\rightarrow l\phi)\;,\label{nbarl}
\eea
with the reaction rates
\bea
\g(N\rightarrow l\phi) &=& \int d\Phi_{123} 
      f_N(p_1) |\cm(N\rightarrow l\phi)|^2\;, \ldots \label{nlphi} \\
\g(l\phi\rightarrow \bar{l}\bar{\phi}) &=& \int d\Phi_{1234}
 f_l(p_1)f_{\phi}(p_2) 
 |\cm'(l\phi\rightarrow \bar{l}\bar{\phi})|^2\;,\ldots\label{llphi}
\eea
Here $H$ is the Hubble parameter, $d\Phi_{1...n}$ denotes the phase space 
integration over particles in initial and final states,
\beq
d\Phi_{1...n}={d^3p_1\over (2\pi)^3 2E_1}\ldots {d^3p_n\over (2\pi)^3 2E_n}
(2\pi)^4 \d^4(p_1+\ldots -p_n)\;,
\eeq
and 
\beq
f_i(p)=\exp{(-\b E_i(p))}\;, \quad 
n_i(p) = g_i \int {d^3p\over (2\pi)^3} f_i(p)\;,
\eeq
are Boltzmann distribution and number density of particle $i=N,l,\phi$ at
temperature $T=1/\b$, respectively. $\cm$ and $\cm'$ denote the scattering 
matrix elements of the indicated processes at zero temperature; the prime
indicates that for the $2\rightarrow 2$ processes the contribution of the 
intermediate resonance state has been subtracted. For simplicity we have used 
in eqs.~(\ref{nlphi}) and
(\ref{llphi}) Boltzmann distributions rather than Bose-Einstein and
Fermi-Dirac distributions, and we have also neglected the distribution
functions in the final state, which is a good approximation for small
number densities. Subtracting (\ref{nbarl}) from (\ref{nl}) yields the
Boltzmann equation for the asymmetry $n_l - n_{\overline{l}}$. 

\begin{figure}[t]
\begin{center}
\epsfig{file=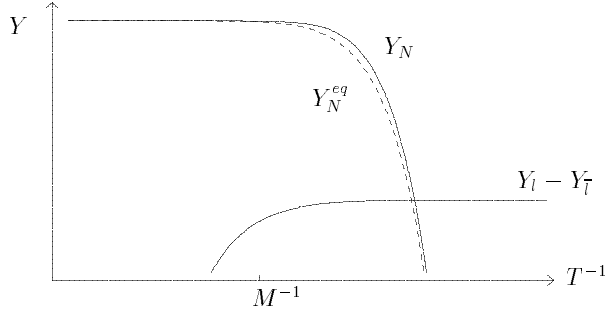,width=8cm,clip=}
\caption{\it Time evolution of the number density to 
entropy density ratio. At $T\sim M$ the system gets out of
equilibrium and an asymmetry is produced.}
\label{fig:outofequ}
\end{center}
\end{figure}

A typical solution of the Boltzmann equations (\ref{nN})~-~(\ref{nbarl}) is
shown in fig.~\ref{fig:outofequ}. Here the ratios of number densities and
entropy density,
\beq
Y_X = {n_X\over s}\;,
\eeq
are plotted, which remain constant in an expanding universe in thermal 
equilibrium. A heavy neutrino, which is weakly coupled 
to the thermal bath, falls out of thermal equilibrium at temperatures 
$T \sim M$ since its decay is too slow to follow the rapidly decreasing 
equilibrium distribution $f_N \sim \exp(-\b M)$. This leads to an excess of 
the number density, $n_N > n_N^{eq}$. $C\!P$ violating partial decay widths then yield a lepton
asymmetry which, by means of sphaleron processes, is partially transformed
into a baryon asymmetry.

The Boltzmann equations are classical equations for the time evolution of
number densities. The collision terms, however, are $S$-matrix elements which
involve quantum mechanical interferences of different amplitudes in a
crucial manner. Since these scattering matrix elements are evaluated at
zero temperature, one may worry to what extent the quantum mechanical
interferences are affected by interactions with the thermal bath. A full
quantum mechanical treatment can be based either on the density matrix
\cite{jmy98} or on on the Kadanoff-Baym equations \cite{bf00}. 

Another subtlety is the separation of the $2\rightarrow 2$ matrix elements 
into a resonance contribution and remainder \cite{kw80},
\beq\label{split}
|\cm(l\phi\rightarrow \bar{l}\bar{\phi})|^2 = 
|\cm'(l\phi\rightarrow \bar{l}\bar{\phi})|^2 + 
|\cm_{res}(l\phi\rightarrow \bar{l}\bar{\phi})|^2\;,
\eeq
where the resonance contribution has the form
\beq
\cm_{res}(l\phi\rightarrow \bar{l}\bar{\phi}) \propto
\cm(l\phi\rightarrow N)\cm(N\rightarrow \bar{l}\bar{\phi})^* =
|\cm(l\phi\rightarrow N)|^2\;.
\eeq 
The entire effect of baryon number generation crucially depends on this
separation. The particles which participate in the $2\rightarrow 2$ processes
are massless, hence their distribution functions always coincide with the
equilibrium distribution. Only the resonances, treated as on-shell particles,
fall out of thermal equilibrium and can therefore generate an asymmetry in 
their decays. General theoretical arguments require cancellations between these
two types of contributions which is illustrated by the following example.

\subsection{Cancellations in thermal equilibrium} 

If all processes, including those which violate baryon number, are in
thermal equilibrium the baryon asymmetry vanishes. This is a direct
consequence of the $C\!PT$ invariance of the theory,
\bea
\langle B \rangle &=& \mbox{Tr}(\r B) 
=\mbox{Tr}\left((C\!PT)(C\!PT)^{-1}\exp{(-\b H)}B\right) \NO\\
&=&\mbox{Tr}\left(\exp{(-\b H)}(C\!PT)^{-1}B(C\!PT)\right) = -\mbox{Tr}(\r B)=0\;.
\eea
Hence, no asymmetry can be generated in equilibrium, and the
transition rate which determines the change of the asymmetry has to vanish,
\beq
{d(n_l-n_{\bar{l}})\over dt} + 3 H(n_l-n_{\bar{l}}) \equiv \D\g^{eq} = 0\;,
\eeq
where the superscript $eq$ denotes rates evaluated with equilibrium 
distributions.

From eqs.~(\ref{gcp}), (\ref{nl}) and (\ref{nbarl}) one obtains for the 
resonance contribution, i.e. decay and inverse decay,
\beq
\D\g^{eq}_{res}=-2\e\g^{eq}(N\rightarrow l\phi)\;.
\eeq
This means in particular that the asymmetry generated in the decay is not
compensated by the effect of inverse decays. On the contrary, both processes
contribute the same amount.

The rate $\D\g^{eq}_{res}$ has to be compensated by the contribution from
$2\rightarrow 2$ processes which is given by
\beq
\D\g^{eq}_{2\rightarrow 2} = 2\int d\Phi_{1234}f_l^{eq}(p_1)f_{\phi}^{eq}(p_2)
\left(|\cm'(l\phi\rightarrow \overline{l}\overline{\phi})|^2 -
      |\cm'(\overline{l}\overline{\phi}\rightarrow l\phi)|^2\right)\;.
\eeq 
For weakly coupled heavy neutrinos, i.e. $\G \propto \l^2 M$ with $\l^2\ll 1$,
this compensation can be easily shown using the unitarity of the $S$-matrix.

The sum over states in the unitarity relation,
\beq
\sum_X\left(|\cm(l\phi\rightarrow X)|^2-|\cm(X\rightarrow l\phi)|^2\right)=0\;,
\eeq
can be restricted to two-particle states to leading order in the case of weak 
coupling $\l$. This implies for the considered $2\rightarrow 2$ processes,
\beq\label{sum}
{\sum_{l,\phi,\bar{l'},\bar{\phi'}}\!\!\!}'
\left(|\cm(l\phi\rightarrow \bar{l'}\bar{\phi'})|^2 -
      |\cm(\bar{l'}\bar{\phi'}\rightarrow l\phi)|^2 \right) = 0\;,
\eeq
where the summation $\sum'$ includes momentum integrations under the 
constraint of fixed total momentum. From eqs.~(\ref{split}) and (\ref{sum})
one obtains        
\bea
\D\g^{eq}_{2\rightarrow 2} &=& 
2\int d\Phi_{1234}f_l^{eq}(p_1)f_{\phi}^{eq}(p_2) \NO\\
&& \hspace{1cm}\left(-|\cm_{res}(l\phi\rightarrow \bar{l}\bar{\phi})|^2 +
      |\cm_{res}(\bar{l}\bar{\phi}\rightarrow l\phi)|^2\right)\;.
\eea
In the narrow width approximation, i.e. to leading order in $\l^2$, this
yields the wanted result,
\bea
\D\g^{eq}_{2\rightarrow 2} &=& 
2\int d\Phi_{1234}f_l^{eq}(p_1)f_{\phi}^{eq}(p_2)
\left(-|\cm(l\phi\rightarrow N)|^2 |\cm(N\rightarrow \bar{l}\bar{\phi})|^2
\right.\NO\\
&&\left. \hspace{2cm}
+|\cm(\bar{l}\bar{\phi}\rightarrow N)|^2|\cm(N\rightarrow l\phi)|^2 \right)
{\pi\over M \G}\d(s-M^2) \NO\\
&=&2 \e \g^{eq}(N\rightarrow l\phi) = - \D\g^{eq}_{res}\;.
\eea
This cancellation also illustrates that the Boltzmann equations treat 
resonances as on-shell real particles. Off-shell effects can only be
taken into account in a full quantum mechanical treatment.

\section{Cosmological bounds on neutrino masses}

Leptogenesis requires lepton number violation. On the other hand, the 
existence of a non-vanishing baryon asymmetry also restricts the allowed 
amount of lepton number violation and implies upper bounds on the Majorana 
masses of the light neutrinos.

In the standard model (SM) neutrinos are massless. In general, however,
the exchange of heavy particles gives rise to an effective lepton Higgs 
interaction which, after electroweak symmetry breaking, generates neutrino 
masses. 
The lagrangian describing all fermion-Higgs couplings then reads 
\bea\label{inter}
\cl_Y &=& - h_{d ij}^T\Bar{d_R}_i q_{Lj} H_1
        - h_{u ij}^T\Bar{u_R}_i q_{Lj} H_2
        - h_{e ij}\Bar{e_R}_i l_{Lj} H_1 \NO\\
&&\hspace{1cm} + {1\over 2} f_{ij}\ l^T_{Li}H_2\ C\ l_{Lj}H_2 +\mbox{ h.c.}\;.
\eea
Here $q_{Li}$, $u_{Ri}$, $d_{Ri}$, $l_{Li}$, $e_{Ri}$, $i=1\ldots N$, are N 
generations of quark and lepton fields, $H_1$ and $H_2$ are Higgs fields with 
vacuum expectation values $v_i=\VEV{H_i^0}\ne0$. $h_d$, $h_u$, $h_e$ and $f$ 
are N$\times$N complex matrices. For the further discussion it is convenient 
to choose a basis where $h_u$ and $h_e$ are diagonal and real. In the case
$v_1 \sim v_2 \sim v=\sqrt{v_1^2+v_2^2}$ the smallest Yukawa coupling is 
$h_{e11}=m_e/v_1$ followed by $h_{u11}=m_u/v_2$. 

The mixing of the different quark generations is given by the 
Kobayashi-Maskawa matrix $V_d$, which is defined by
\beq
 h_d^T\ V_d\ = \ {m_d\over v_1}\;.
\eeq
Here $m_d$ is the diagonal real down quark mass matrix, and the weak 
eigenstates of the right-handed d-quarks have been chosen to be identical to
the mass eigenstates. Correspondingly,
the mixing matrix $V_\n$ in the leptonic charged current is determined by
\beq
 V_\n^T\ f\ V_\n\ =\ -\ {m_\n \over v_2^2}\;, 
 \label{vnu}
\eeq
where 
\beq
m_\n=\left(\begin{array}{ccc}m_1&0&0\\0&m_2&0\\0&0&m_3
              \end{array}\right)
\eeq 
is the diagonal and real mass matrix of the light Majorana neutrinos.

\subsection{Chemical equilibrium}
  
In a weakly coupled plasma with temperature $T$ and volume $V$ one can
assign a chemical potential $\m$ to each of the quark, lepton and Higgs
fields. In the SM with one Higgs doublet, i.e., $H_2=\wt H_1 \equiv \vf$,
and N generations one has 5N+1 chemical potentials. The corresponding 
partition function is \cite{lan}
\beq
Z(\m,T,V) = \mbox{Tr}e^{-\b(H-\sum_i\m_i Q_i)}\;.
\eeq
Here $\b=1/T$, $H$ is the Hamilton operator and $Q_i$ are the charge 
operators for the corresponding fields. The asymmetry in the particle and
antiparticle number densities is then given by the derivative of the 
thermodynamic potential,
\beq
n_i-\overline{n}_i=-{\partial \O(\m,T)\over \partial \m_i}\;, \quad
\O(\m,T) = - {T\over V} \ln{Z(\m,T,V)}\;.
\eeq
For a non-interacting gas of massless particles one has
\beq\label{number}
n_i-\overline{n}_i={g T^3\over 6}
\left\{\begin{array}{rl}\b\m_i +{\cal O}\left(\left(\b\m_i\right)^3\right)\;,
&\mbox{fermions}\;,\\
2\b\m_i+{\cal O}\left(\left(\b\m_i\right)^3\right)\;, &\mbox{bosons}\;.
\end{array}\right.
\eeq
The following analysis will be based on these relations for $\b \m_i \ll 1$.
However, one should keep in mind that the plasma of the early universe is
very different from a weakly coupled relativistic gas due to the presence
of unscreened non-abelian gauge interactions. Hence, non-perturbative effects
may be important in some cases.

In the high-temperature plasma quarks, leptons and Higgs bosons interact via
Yukawa and gauge couplings and, in addition, via the non-perturbative
sphaleron processes. In thermal equilibrium all these processes yield
constraints between the various chemical potentials. The effective interaction
(\ref{obl}) induced by the $SU(2)$ electroweak instantons yields the 
constraint \cite{krs85},
\beq\label{sphew}
\sum_i\left(3\m_{qi} + \m_{li}\right) = 0\;.
\eeq
One also has to take the $SU(3)$ QCD instanton processes into account 
\cite{mz92} which generate the effective interaction
\beq\label{oaxi}
O_A = \prod_i \left(q_{Li}q_{Li}u^c_{Ri}d^c_{Ri}\right)
\eeq
between left-handed and right-handed quarks. The corresponding relation
between the chemical potentials reads
\beq\label{sphqcd}
\sum_i\left(2\m_{qi} - \m_{ui} - \m_{di}\right) = 0\;.
\eeq
A third condition, which is valid at all temperatures, arises from the
requirement that the total hypercharge of the plasma vanishes. From
eq.~(\ref{number}) and the known hypercharges one obtains
\beq\label{hypsm}
\sum_i\left(\m_{qi} + 2 \m_{ui} - \m_{di} - \m_{li} - \m_{ei} + 
{2\over N} \m_{\vf}\right) = 0\;.
\eeq

The Yukawa interactions, supplemented by gauge interactions, yield relations
between the chemical potentials of left-handed and right-handed fermions,
\beq\label{myuk}
\m_{qi}-\m_{\vf}-\m_{dj} = 0\;, \quad
\m_{qi}+\m_{\vf}-\m_{uj} = 0\;, \quad
\m_{li}-\m_{\vf}-\m_{ej} = 0\;.
\eeq
Furthermore, the $\D L =2$ interaction in (\ref{inter}) implies
\beq\label{mdl2}
\m_{li}+\m_{\vf} = 0\;.
\eeq

The above relations between chemical potentials hold if the corresponding
interactions are in thermal equilibrium. In the temperature range
$T_{EW} \sim 100\ \mbox{GeV} < T < T_{SPH} \sim 10^{12}\ \mbox{GeV}$, 
which is of interest for 
baryogenesis, this is the case for all gauge interactions. It is not always 
true, however, for Yukawa interactions. The rate of a scattering process 
between left- and right-handed fermions, Higgs boson and W-boson, 
\beq
\psi_L \vf \rightarrow \psi_R W\;,
\eeq
is $\G \sim \a \l^2 T$, with $\a=g^2/(4\pi)$. This rate has to be
compared with the Hubble rate,
\beq
H \simeq 0.33\ g_*^{1/2} {T^2\over M_{PL}} 
  \simeq 0.1\ g_*^{1/2} {T^2\over 10^{18}\ \mbox{GeV}}\;.
\eeq  
The equilibrium condition $\G(T) > H(T)$ is satisfied for sufficiently small 
temperatures, 
\beq\label{eqyuk}
T < T_\l \sim \l^2 10^{16}\ \mbox{GeV}\; .
\eeq
Hence, one obtains for the decoupling temperatures of right-handed electrons,
up-quarks,...,
\beq
T_e \sim 10^4\ \mbox{GeV}\;,\quad T_u \sim 10^6\ \mbox{GeV}\;,\ldots .
\eeq
At a temperature $T \sim 10^{10}$ GeV, which is characteristic of leptogenesis,
$e_R\equiv e_{R1}$, $\m_R\equiv e_{R2}$, $d_R\equiv d_{R1}$, $s_R\equiv d_{R2}$
and $u_R\equiv u_{R1}$ are out of equilibrium.

Note that the QCD sphaleron constraint (\ref{sphqcd}) is automatically
satisfied if the quark Yukawa interactions are in equilibrium 
(cf.~(\ref{myuk})). If the Yukawa interaction of one of the right-handed 
quarks is too weak, the sphaleron constraint still establishes full
chemical equilibrium. 

Using eq.~(\ref{number}) also the baryon number density $n_B \equiv B T^2/6$
and the lepton number densities $n_L \equiv L T^2/6$ can be expressed in terms
of the chemical potentials. The baryon asymmetry $B$ and the lepton asymmetries
$L_i$ read
\bea
B &=& \sum_i \left(2\m_{qi} + \m_{ui} + \m_{di}\right)\;, \\
L_i &=& 2\m_{li} + \m_{ei}\;,\quad L=\sum_i L_i\;.
\eea

\subsection{Relations between $B$, $L$ and $B-L$}

Knowing which particle species are in thermal equilibrium one can derive
relations between different asymmetries. Consider first the most familiar
case where all Yukawa interactions are in equilibrium and the $\D L=2$
lepton-Higgs interaction is out of equilibrium. In this case the asymmetries 
$L_i-B/N$ are conserved. The Yukawa interactions establish equilibrium 
between the different generations,
\beq
\m_{li} \equiv \m_l\;, \quad \m_{qi} \equiv \m_q\;, \quad \mbox{etc.}
\eeq  
Together with the sphaleron process and the hypercharge constraint they
allow to express all chemical potentials, and therefore all asymmetries,
in terms of a single chemical potential which may be chosen to be $\m_l$.
The result reads
\bea\label{exam1}
\m_e &=& {2N+3\over 6N+3}\m_l\;, \quad
\m_d = -{6N+1\over 6N+3}\m_l\;, \quad
\m_u = {2N-1\over 6N+3}\m_l\;, \NO\\
\m_q &=& -{1\over 3} \m_l\;, \quad
\m_{\vf} = {4N\over 6N+3} \m_l\;.
\eea
The corresponding baryon and lepton asymmetries are
\beq
B = -{4N\over 3}\m_l\;, \quad L = {14N^2+9N\over 6N+3}\m_l\;,
\eeq
which yields the well-known connection between the $B$ and $B-L$ 
asymmetries \cite{ks88}
\beq
B = {8N+4\over 22N+13} (B-L)\;.
\eeq
Note, that this relation only holds for temperatures $T\gg v$. In general,
the ratio $B/(B-L)$ is a function of $v/T$ \cite{ks96,ls00}.

Another instructive example is the case where the $\D L=2$ interactions are
in equilibrium but the right-handed electrons are not. Depending on the 
neutrino masses and mixings, this could be the case for temperatures above
$T_e \sim 10^4$ GeV \cite{cko93}. Right-handed electron number would then be
conserved, and Yukawa and gauge interactions would relate all asymmetries
to the asymmetry of right-handed electrons. The various chemical potentials
are given by ($\m_e = \m_{e1}$, $\m_{\bar{e}} = \m_{e2}=\ldots=\m_{eN}$),
\bea\label{exam2}
\m_{\bar{e}}&=&-{3\over 10 N}\m_e\;,\quad 
\m_d = -{1\over 10 N}\m_e\;,\quad
\m_u = {1\over 5 N}\m_e\;, \NO\\
\m_l&=&-{3\over 20 N}\m_e\;,\quad
\m_q = {1\over 20 N}\m_e\;, \quad
\m_{\vf} = {3\over 20 N}\m_e\;.
\eea
The corresponding baryon and lepton asymmetries are \cite{cko93}
\beq
B = {1\over 5}\m_e \;, \quad L = {4N+3\over 10 N}\m_e\;,
\eeq
which yields for the relation between $B$ and $B-L$,
\beq
B = -{2N\over 2N+3} (B-L)\;.
\eeq
Note that although sphaleron processes and $\D L=2$ processes are in
equilibrium, the asymmetries in $B$, $L$ and $B-L$ do not vanish! 

\subsection{Constraint on Majorana neutrino masses}

The two examples illustrate the connection between lepton number and
baryon number induced by sphaleron processes. They also show how this
connection depends on other processes in the high-temperature plasma.
To have one quark-Higgs or lepton-Higgs interaction out of equilibrium
is sufficient in order to have non-vanishing $B$, $L$ and $B-L$. If
all interactions in (\ref{inter}) are in equilibrium, eqs.~(\ref{mdl2})
and (\ref{exam1}) together imply $\m_l=0$ and therefore
\beq
B = L = B-L = 0\;,
\eeq
which is inconsistent with the existence of a matter-antimatter asymmetry.
Since the equilibrium conditions of the various interactions are
temperature dependent, and the $\D L=2$ interaction is related to neutrino
masses and mixings, one obtains important constraints on neutrino properties
from the existence of the cosmological baryon asymmetry.

The $\D L=2$ processes described by (\ref{dl2}) take place with the 
rate \cite{fy90}
  \beq
    \Gamma_{\Delta L=2} (T) = {1\over \pi^3}\,{T^3\over v^4}\, 
    \sum_{i=e,\m,\t} m_{\n_i}^2\; .
  \eeq
Requiring $\G_{\D L=2}(T) < H(T)$ then yields an upper bound on Majorana 
neutrino masses,
\beq\label{nbound}
\sum_i m_{\n_i}^2 < \left(0.2\ \mbox{eV}\ \left({T_{SPH}\over T}
                    \right)^{1/2} \right)^2\;.
\eeq
For typical leptogenesis temperatures $T \sim 10^{10}$~GeV this bound is 
comparable to the upper bound on the electron neutrino mass 
obtained from neutrinoless double beta decay. Note, that the bound
also applies to the $\t$-neutrino mass. However, if one uses for $T$ the
decoupling temperature of right-handed electrons, $T_e \sim 10^4$~GeV, the
much weaker bound $m_\n < 2$~keV is obtained \cite{cko93}.

Clearly, what temperature one has to use in eq.~(\ref{nbound}) depends on
the thermal history of the early universe. Some information is needed on
what kind of asymmetries may have been generated as the temperature decreased.
This, together with the temperature dependence of the lepton-Higgs interactions
then yields constraints on neutrino masses. 

\subsection{Primordial asymmetries}

The possible generation of asymmetries can be systematically studied by
listing all the higher-dimensional $SU(3)\times SU(2)\times U(1)$
operators which may be generated by
the exchange of heavy particles. The dynamics of the heavy particles may
then generate an asymmetry in the quantum numbers carried by the massless
fields which appear in the operator.

For d=5, there is a unique operator, which has already been discussed
above,
\beq
(l_{Li}\vf)(l_{Lj}\vf)\;.
\eeq
It is generated in particular by the exchange of heavy Majorana neutrinos
whose coupling to the massless fields is
\beq
h_{\n ij}\Bar{\n_R}_i l_{Lj} \vf \;.
\eeq
The out-of-equilibrium decays of the heavy neutrinos can generate a lepton 
asymmetry, which is the well-known mechanism of leptogenesis. The decays
yield asymmetries $L_i-B/N$ which are conserved in the subsequent evolution.  
The initial asymmetry in right-handed electrons is zero. In order to satisfy 
the out-of-equilibrium condition it is very important that
at least some Yukawa couplings are small and that the right-handed neutrinos
carry no quantum numbers with respect to unbroken gauge symmetries.

In order to study possible asymmetries of right-handed electrons one has
to consider operators containing $e_R$. A simple example, with d=6, reads
\beq
(q_{Li}l_{Lj})(u^c_{Rk}e^c_{Rl})\;.
\eeq
It can be generated by leptoquark exchange ($\chi \sim (3^*,1,1/3)$),
\beq
(\l^q_{ij} q^T_{Li}C l_{Lj} + \l^u_{ij} u^T_{Ri}C e_{Rj})\chi\;.
\eeq
Note, that $\wt{\vf}$ and $\chi$ form a $5^*$-plet of $SU(5)$. In principle,
out-of-equilibrium decays of leptoquarks may generate a $e_R$ asymmetry.
One may worry, however, whether the branching ratio into final states
containing $e_R$ is sufficiently large. Furthermore, it appears very
difficult to satisfy the out-of-equilibrium condition since leptoquarks
carry colour. Maybe, all these problems can be overcome by making use of
coherent oscillations of scalar fields \cite{ad85} or by special particle 
production mechanisms after inflation. However, we are not aware of a
consistent scenario for the generation of a $e_R$-asymmetry. 
Hence, it appears appropriate to take the bound
eq.~(\ref{nbound}) on Majorana neutrino masses as a guideline and to examine
its validity in each particular model.

\section{Neutrino masses and mixings}

Majorana masses for the light neutrinos are most easily generated by the
exchange of heavy Majorana neutrinos. Such heavy `right-handed' neutrinos
are predicted by all extensions of the standard model which contain $B-L$
as a local symmetry. The most general Lagrangian for couplings and masses of 
charged leptons and neutrinos reads
  \beq\label{yuk}
  \cl_Y = -h_{e ij}\Bar{e_R}_i l_{Lj} H_1 
          -h_{\n ij}\Bar{\n_R}_i l_{Lj} H_2
          -{1\over2}h_{r ij} \Bar{\n^c_R}_i \n_{Rj} R +\mbox{ h.c.}\;.
  \eeq
  The vacuum expectation values of the Higgs fields, $\VEV{H_1}=v_1$ and
  $\VEV{H_2}=v_2=\tan{\b}\ v_1$, generate Dirac masses $m_e$ and $m_D$
  for charged leptons and neutrinos, $m_e=h_e v_1$ and
  $m_D=h_{\n}v_2$, respectively, which are assumed to be much smaller
  than the Majorana masses $M = h_r\VEV{R}$.  This yields light and
  heavy neutrino mass eigenstates according to the seesaw mechanism
  \cite{seesaw},
  \beq
     \n\simeq V_{\nu}^T\n_L+\n_L^c V_{\nu}^*\quad,\qquad
     N\simeq\n_R+\n_R^c\, ,
  \eeq
  with masses
  \beq
     m_{\n}\simeq- V_{\nu}^Tm_D^T{1\over M}m_D V_{\nu}\,
     \quad,\quad  m_N\simeq M\, .
     \label{seesaw}
  \eeq
  Here $V_{\nu}$ is the mixing matrix in the leptonic charged current
  (cf.\ eqs.~(\ref{inter})-(\ref{vnu})).

In models of leptogenesis the predicted value of the baryon asymmetry
depends on the CP asymmetry (cf.~(\ref{gcp})) which is determined by the 
Dirac and 
the Majorana neutrino mass matrices. Depending on the neutrino mass hierarchy 
and the size of the mixing angles the CP asymmetry can vary over many orders of
magnitude. It is therefore important to see whether patterns of neutrino
masses \cite{EA} motivated by other considerations are consistent with 
leptogenesis. In the following we shall consider two examples.

An attractive framework to explain the observed mass hierarchies of quarks
and charged leptons is the Froggatt-Nielsen mechanism \cite{fn79} based
on a spontaneously broken $U(1)_F$ generation symmetry. The Yukawa couplings 
arise from non-renormalizable interactions after a gauge singlet field $\F$ 
acquires a vacuum expectation value,
\beq
h_{ij} = g_{ij} \left({\VEV\F\over \L}\right)^{Q_i + Q_j}\;.
\eeq
Here $g_{ij}$ are couplings $\co(1)$ and $Q_i$ are the $U(1)_F$ charges of the
various fermions, with $Q_{\F}=-1$. The interaction scale $\L$ is
usually chosen to be very large, $\L > \L_{GUT}$. In the following we shall
discuss two different realizations of this idea which are motivated by
the atmospheric neutrino anomaly \cite{atm98}. Both
scenarios have a large $\n_\m -\n_\t$ mixing angle. They differ, however,
by the symmetry structure and by the size of the parameter $\e$ which
characterizes the flavour mixing.

\subsection{$SU(5)\times U(1)_F$}
 
This symmetry has been considered by a number of authors.
Particularly interesting is the case with a non-parallel family structure
where the chiral $U(1)_F$ charges are different for the $\bf 5^*$-plets
and the $\bf 10$-plets of the same family \cite{ys99}-\cite{by99}. An
example of possible charges $Q_i$ is given in table~1.

\begin{table}[b]
\begin{center}
\begin{tabular}{c|ccccccccc}\hline \hline
$\j_i$       & $ e^c_{R3}$ & $ e^c_{R2}$  & $ e^c_{R1}$  & $ l_{L3}$    & 
$ l_{L2}$    & $ l_{L1}$   & $ \n^c_{R3}$ & $ \n^c_{R2}$ & $ \n^c_{R1}$ 
\\\hline
$Q_i$  & 0 & 1 & 2 & $0$ & $0$ & $1$ & 0 & $1$ & $2$ \\ \hline\hline
\end{tabular}
\end{center}
\caption{{\it Chiral charges of charged and neutral leptons with
   $SU(5)\times U(1)_F$ symmetry} \protect\cite{by99}.}
\end{table}

The assignment of the same charge to the lepton doublets of the second and 
third generation leads to a neutrino mass matrix of the form
\cite{ys99,ram99}, 
\beq\label{matrix}
m_{\n_{ij}} \sim \left(\begin{array}{ccc}
    \e^2  & \e  & \e \\
    \e  & \; 1 \; & 1 \\
    \e  &  1  & 1 
    \end{array}\right) {v_2^2\over \VEV R}\;.
\eeq
This structure immediately yields a large $\n_\m -\n_\t$ mixing angle. The
phenomenology of neutrino oscillations depends on the unspecified coefficients
$\co(1)$. The parameter $\e$ which gives the flavour mixing is chosen to be
\beq\label{exp1}
{\VEV\F\over\L} = \e  \sim {1\over 17}\;, 
\eeq
which corresponds to the mass ratio $m_\m/m_\t$.
The three Yukawa matrices for the leptons are given by,
\beq\label{yuk1}
h_e, \ 
h_{\n}  \sim\ \left(\begin{array}{ccc}
    \e^3 & \e^2 & \e^2 \\
    \e^2 &\;  \e \;   & \e   \\
    \e   & 1    & 1
    \end{array}\right) \;, \quad
h_{r}  \sim\ \left(\begin{array}{ccc}
    \e^4 & \e^3 & \e^2 \\
    \e^3 &\;  \e^2 \; & \e   \\
    \e^2 & \e   & 1
    \end{array}\right) \;.
\eeq
Note, that $h_e$ and $h_\n$ have the same, non-symmetric structure.
One easily verifies that the mass ratios for charged leptons, heavy and
light Majorana neutrinos are given by 
\bea
\qquad\quad 
m_e : m_\m : m_\t \sim \e^3 : \e : 1\;, \quad
M_1 : M_2  : M_3  \sim \e^4 : \e^2 : 1\;,\\
m_1 : m_2  : m_3  \sim \e^2 : 1 : 1\;.
\eea
The masses of the two eigenstates $\n_2$ and $\n_3$ depend on the unspecified 
factors of order one, and may differ by an order of magnitude 
\cite{ilr98,vis98}. They can therefore be consistent with the mass differences 
$\D m^2_{\n_1 \n_2}\simeq 4\cdot 10^{-6} - 1\cdot 10^{-5}$~eV$^2$ \cite{hl97}
inferred from the MSW solution of the solar neutrino problem \cite{msw86} and 
$\D m^2_{\n_2 \n_3}\simeq (5\cdot 10^{-4}-6\cdot 10^{-3})$~eV$^2$ associated 
with the atmospheric neutrino deficit \cite{atm98}. In the following we
we shall use for numerical estimates
the average of the neutrino masses of the second and third family,
$\Bar{m}_\n=(m_{\n_2}m_{\n_3})^{1/2} \sim 10^{-2}$~eV. 

The choice of the charges in table~1 corresponds to large Yukawa couplings
of the third generation. For the mass of the heaviest Majorana neutrino
one finds
\beq
M_3\ \sim\ {v_2^2\over\Bar{m}_\n}\ \sim\ 10^{15}\ \mbox{GeV}\;.
\eeq
Since $h_{r33}$ and the gauge coupling of $U(1)_{B-L}$ are ${\cal O}(1)$,
this implies that $B-L$ is broken at the unification scale $\L_{GUT}$.

\subsection{$SU(3)_c \times SU(3)_L \times SU(3)_R \times U(1)_F$}

This symmetry arises in unified theories based on the gauge group $E_6$.
The leptons $e_R^c$, $l_L$ and $\n_R^c$ are contained in a single
$(1,3,\bar{3})$ representation. Hence, all leptons of the same generation
have the same $U(1)_F$ charge and all leptonic Yukawa matrices are
symmetric. Masses and mixings of quarks and charged leptons can be
successfully described by using the charges given in table~2. 
Clearly, all three Yukawa matrices have the same structure\footnote{Note,
that with respect to ref.~\cite{lr99}, $\e$ and $\Bar{\e}$ have been
interchanged.},
\beq\label{yuk2}
h_e,\ h_r  \sim\ \left(\begin{array}{ccc}
    \e^5        & \e^3        & \e^{5/2} \\
    \e^3        & \e          & \e^{1/2} \\
    \e^{5/2} \; & \e^{1/2} \; & 1
    \end{array}\right) \;, \quad
h_{\n}  \sim\ \left(\begin{array}{ccc}
    \bar\e^5        &\bar\e^3        & \bar\e^{5/2} \\
    \bar\e^3        &\bar\e          & \bar\e^{1/2} \\
    \bar\e^{5/2} \; &\bar\e^{1/2} \; & 1
    \end{array}\right) \;,
\eeq
but the expansion parameter in $h_{\n}$ may be different from the one in
$h_e$ and $h_r$. From the quark masses, which also contain $\e$ and
$\bar{\e}$, one infers $\bar{\e} \simeq \e^2$ \cite{lr99}.

\begin{table}
\begin{center}
\begin{tabular}{c|ccccccccc}
\hline \hline
$\j_i$       & $ e^c_{R3}$ & $ e^c_{R2}$  & $ e^c_{R1}$  & $ l_{L3}$    & 
$ l_{L2}$    & $ l_{L1}$   & $ \n^c_{R3}$ & $ \n^c_{R2}$ & $ \n^c_{R1}$ 
\\\hline
$Q_i$  & 0 & ${1\over 2}$ & ${5\over 2}$ & $0$ & ${1\over 2}$ & ${5\over 2}$ 
& 0 & ${1\over 2}$ & ${5\over 2}$ \\ \hline\hline
\end{tabular}
\end{center}
\caption{{\it Chiral charges of charged and neutral leptons with
$SU(3)_c \times SU(3)_L \times SU(3)_R \times U(1)_F$ symmetry}
\protect\cite{lr99}.}
\end{table}

From eq.~(\ref{yuk2}) one obtains for the masses of charged leptons,
light and heavy Majorana neutrinos,
\beq
m_e : m_\m : m_\t\ \sim\ M_1 : M_2  : M_3\ \sim \e^5 : \e : 1\;, 
\eeq
\beq
m_1 : m_2 : m_3 \ \sim\ \e^{15} : \e^3 : 1\;.
\eeq
Like in the example with $SU(5)\times U(1)_F$ symmetry, the mass of the
heaviest Majorana neutrino,
\beq
M_3 \sim {v_2^2\over m_3} \sim 10^{15}\;\mbox{GeV} \;,
\eeq
implies that $B-L$ is broken at the unification scale $\L_{GUT}$.

The $\n_\m-\n_\t$ mixing angle is related to the mixing of
the charged leptons of the second and third generation \cite{lr99},
\beq
\sin{\Q_{\m\t}} \sim \sqrt{\e} + \e \;.
\eeq
This requires large flavour mixing,
\beq\label{exp2}
\left({\VEV\F\over\L}\right)^{1/2} = \sqrt{\e}  \sim {1\over 2}\;. 
\eeq 
In view of the unknown coefficients $\co(1)$ the corresponding mixing angle 
$\sin{\Q_{\m\t}} \sim 0.7$ is consistent with the interpretation of the 
atmospheric neutrino anomaly as $\n_\m-\n_\t$ oscillation.

It is very instructive to compare the two scenarios of lepton masses and
mixings described above. In the first case, the large $\n_\m-\n_\t$
mixing angle follows from a non-parallel flavour symmetry. The parameter $\e$,
which characterizes the flavour mixing, is small. In the second case, the
large $\n_\m-\n_\t$ mixing angle is a consequence of the large flavour
mixing $\e$. The $U(1)_F$ charges of all leptons are the same, i.e., one
has a parallel family structure. Also the mass hierarchies, given in terms
of $\e$, are rather different. This illustrates that the separation into
a flavour mixing parameter $\e$ and coefficients $\co(1)$ is far from
unique. It is therefore important to study other observables which
depend on the lepton mass matrices. This includes lepton flavour 
changing processes and, in particular, the cosmological baryon 
asymmetry.

\section{Calculating the baryon asymmetry}

\subsection{Majorana neutrino decays}
  
  The heavy Majorana neutrinos, whose exchange may erase any lepton
  asymmetry, can also generate a lepton asymmetry by means of
  out-of-equilibrium decays. This lepton asymmetry is then partially 
  transformed into a baryon asymmetry by sphaleron processes.  
  The decay width of the heavy neutrino $N_i$ reads at tree level,
  \beq
    \G_{Di}=\G\left(N_i\to H_2+l\right)+\G\left(N_i\to H_2^c+l^c\right)
           ={1\over8\p}(h_\n h_\n^\dg)_{ii} M_i\;.
    \label{decay}
  \eeq
A necessary requirement for baryogenesis is the out-of-equilibrium condition
$\Gamma_{D1}< H|_{T=M_1}$ \cite{kt90}, where $H$ is the Hubble parameter
at temperature $T$. From the decay width (\ref{decay}) one then obtains an 
upper bound on an effective light neutrino mass \cite{fgx91,by93}, 
\bea
    \wt{m}_1\ &=&\ (h_\n h_\n^\dg)_{11} {v_2^2\over M_1}\ 
    \simeq 4 g_*^{1/2} {v_2^2\over M_P}
    \left.{\G_{D1}\over H}\right|_{T=M_1}\NO\\
    &<& 10^{-3}\, \mbox{eV}\;.\label{ooeb}
\eea
Here $g_*$ is the number of relativistic degrees of freedom, 
$M_P=(8\pi G_N)^{-1/2}\simeq 2.4\cdot 10^{18}$~GeV is the Planck mass, and 
we have assumed $g_*\simeq 100$, $v_2\simeq 174$~GeV.
  More direct bounds on the light neutrino masses depend on the structure
  of the Dirac neutrino mass matrix. If the bound (\ref{ooeb}) is satisfied,
  the heavy neutrinos $N_1$ are not able to follow the rapid change of the
  equilibrium distribution once the temperature of the
  universe drops below the mass $M_1$. Hence, the deviation from
  thermal equilibrium consists in a too large number density of heavy 
neutrinos $N_1$ as compared to the equilibrium density (cf.~section~2.1). 

  \begin{figure}[t]
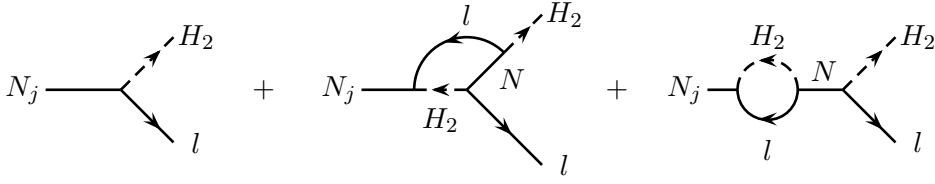

    \centerline{\parbox[c]{12.5cm}{
\pspicture(0,0)(3.7,2.6)
\psline[linewidth=1pt](0.6,1.3)(1.6,1.3)
\psline[linewidth=1pt](1.6,1.3)(2.3,0.6)
\psline[linewidth=1pt,linestyle=dashed](1.6,1.3)(2.3,2.0)
\psline[linewidth=1pt]{->}(2.03,0.87)(2.13,0.77)
\psline[linewidth=1pt]{->}(2.03,1.73)(2.13,1.83)
\rput[cc]{0}(0.3,1.3){$N_j$}
\rput[cc]{0}(2.6,0.6){$l$}
\rput[cc]{0}(2.6,2.0){$H_2$}
\rput[cc]{0}(3.5,1.3){$+$}
\endpspicture
\pspicture(-0.5,0)(4.2,2.6)
\psline[linewidth=1pt](0.6,1.3)(1.3,1.3)
\psline[linewidth=1pt,linestyle=dashed](1.3,1.3)(2.0,1.3)
\psline[linewidth=1pt](2,1.3)(2.5,1.8)
\psline[linewidth=1pt,linestyle=dashed](2.5,1.8)(3,2.3)
\psline[linewidth=1pt](2,1.3)(3,0.3)
\psarc[linewidth=1pt](2,1.3){0.7}{45}{180}
\psline[linewidth=1pt]{<-}(1.53,1.3)(1.63,1.3)
\psline[linewidth=1pt]{<-}(1.7,1.93)(1.8,1.96)
\psline[linewidth=1pt]{->}(2.75,2.05)(2.85,2.15)
\psline[linewidth=1pt]{->}(2.5,0.8)(2.6,0.7)
\rput[cc]{0}(0.3,1.3){$N_j$}
\rput[cc]{0}(1.65,0.9){$H_2$}
\rput[cc]{0}(2,2.3){$l$}
\rput[cc]{0}(2.6,1.45){$N$}
\rput[cc]{0}(3.3,2.3){$H_2$}
\rput[cc]{0}(3.3,0.3){$l$}
\rput[cc]{0}(4.0,1.3){$+$}
\endpspicture
\pspicture(-0.5,0)(3.5,2.6)
\psline[linewidth=1pt](0.5,1.3)(0.9,1.3)
\psline[linewidth=1pt](1.7,1.3)(2.3,1.3)
\psarc[linewidth=1pt](1.3,1.3){0.4}{-180}{0}
\psarc[linewidth=1pt,linestyle=dashed](1.3,1.3){0.4}{0}{180}
\psline[linewidth=1pt]{<-}(1.18,1.69)(1.28,1.69)
\psline[linewidth=1pt]{<-}(1.18,0.91)(1.28,0.91)
\psline[linewidth=1pt](2.3,1.3)(3.0,0.6)
\psline[linewidth=1pt,linestyle=dashed](2.3,1.3)(3.0,2.0)
\psline[linewidth=1pt]{->}(2.73,0.87)(2.83,0.77)
\psline[linewidth=1pt]{->}(2.73,1.73)(2.83,1.83)
\rput[cc]{0}(1.3,0.5){$l$}
\rput[cc]{0}(1.3,2){$H_2$}
\rput[cc]{0}(2.05,1.55){$N$}
\rput[cc]{0}(0.2,1.3){$N_j$}
\rput[cc]{0}(3.3,0.6){$l$}
\rput[cc]{0}(3.3,2.0){$H_2$}
\endpspicture
}}
    \caption{\it Tree level and one-loop diagrams contributing to heavy
    neutrino decays. \label{decay_fig}}
  \end{figure}

  Eventually, however, the neutrinos will decay, and a lepton
  asymmetry is generated due to the CP asymmetry which comes about
  through interference between the tree-level amplitude and the
  one-loop diagrams shown in fig.~\ref{decay_fig}.  In a basis, where
  the right-handed neutrino mass matrix $M = h_r\VEV{R}$ is diagonal,
  one obtains
  \bea
    \ve_1&=&
    {\Gamma(N_1\rightarrow l \, H_2)-\Gamma(N_1\rightarrow l^c \, H_2^c)
    \over
    \Gamma(N_1\rightarrow l \, H_2)+\Gamma(N_1\rightarrow l^c \, H_2^c)}
    \NO\\[1ex]    
    &\simeq&{1\over8\pi}\;{1\over\left(h_\n h_\n^\dg\right)_{11}}
    \sum_{i=2,3}\mbox{Im}\left[\left(h_\n h_\n^\dg\right)_{1i}^2\right]
    \left[f\left(M_i^2\over M_1^2\right)+
    g\left(M_i^2\over M_1^2\right)\right]\label{cpa}\; ;
  \eea
  here $f$ is the contribution from the one-loop vertex correction, 
  \beq
    f(x)=\sqrt{x}\left[1-(1+x)\ln\left(1+x\over x\right)\right]\;,
  \eeq
  and $g$ denotes the contribution from the one-loop self energy
  \cite{fps95,crv96,bp98}, which can be reliably calculated in perturbation
  theory for sufficiently large mass splittings, i.e.,
  $|M_i-M_1|\gg|\G_i-\G_1|$,
  \beq
    g(x)={\sqrt{x}\over1-x}\;.
  \eeq
For $M_1 \ll M_2, M_3$ one obtains
\beq\label{eps}
    \ve_1 \simeq -{3\over16\pi}\;{1\over\left(h_\n h_\n^\dg\right)_{11}}
    \sum_{i=2,3}\mbox{Im}\left[\left(h_\n h_\n^\dg\right)_{1i}^2\right]
    {M_1\over M_i}\; .
  \eeq
In the case of mass differences of order the decay widths 
one expects an enhancement from the self-energy contribution \cite{pil99}.

  The CP asymmetry (\ref{cpa}) leads to a lepton
  asymmetry \cite{kt90},
  \beq\label{basym}
    Y_L\ =\ {n_L-n_{\Bar{L}}\over s}\ =\ \k\ {\ve_1\over g_*}\;.
  \eeq
  Here the factor $\k<1$ represents the effect of washout
  processes. In order to determine $\k$ one has to solve the full
  Boltzmann equations \cite{lut92,plu97}. In the examples discussed
  in sections 4.1 and 4.2 one obtains $\k\simeq 10^{-1}\ldots 10^{-3}$.
  Important processes are the $\D L=2$ lepton Higgs scatterings
  mediated by heavy neutrinos (cf.~fig.~5) since
  cancellations between on-shell contributions to these scatterings
  and contributions from neutrino decays and inverse decays ensure
  that no asymmetry is generated in thermal equilibrium \cite{kw80}.
  Further, due to the large top-quark Yukawa coupling one has to take
  into account neutrino top-quark scatterings mediated by Higgs bosons
  \cite{lut92,plu97}. 
  These processes are of crucial importance for leptogenesis, since
  they can create a thermal population of heavy neutrinos at high
  temperatures $T>M_1$. Alternatively, the density of heavy neutrinos
  may be generated by inflaton decays \cite{GL}. Obviously,
  the requested baryon asymmetry can only be generated if
  the heavy neutrinos are sufficiently numerous before decaying.

Leptogenesis has been considered for various extensions of the standard model
(cf.~\cite{bp00}). In particular, it is
intriguing that in the simple case of hierarchical heavy neutrino masses
the observed value of the baryon asymmetry is obtained without any fine tuning
of parameters if $B-L$ is broken at the unification scale, 
$\L_{GUT} \sim 10^{16}$~GeV \cite{bp96}. The corresponding light neutrino 
masses are very small, i.e., 
$m_{\n_2} \sim 3\cdot 10^{-3}$~eV, as preferred by the MSW explanation of
the solar neutrino deficit, and $m_{\n_3} \sim 0.1$~eV. 
Such small neutrino masses are also consistent with the atmospheric neutrino
anomaly \cite{atm98}, which implies a small mass $m_{\n_3}$ in the
case of hierarchical neutrino masses. This fact  gave rise to a renewed 
interest in leptogenesis in recent years.

\subsection{Baryon asymmetry and baryogenesis temperature}

We can now evaluate the baryon asymmetry for the two patterns of neutrino
mass matrices discussed in sections 4.1 and 4.2.
Since for the Yukawa couplings only the powers in $\e$ are known, we will
also obtain the CP asymmetries and the corresponding baryon asymmetries
to leading order in $\e$, i.e., up to unknown factors ${\cal O}(1)$. 
Note, that for models with a $U(1)_F$ generation symmetry the baryon
asymmetry is `quantized', i.e., changing the $U(1)_F$ charges will change
the baryon asymmetry by powers of $\e$ \cite{by99}.

\subsubsection{$SU(5)\times U(1)_F$}

\begin{figure}[t]
    \mbox{ }\hfill
    \epsfig{file=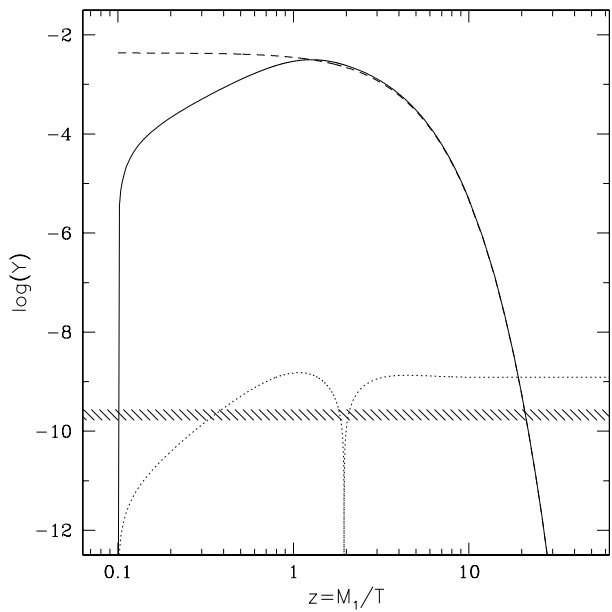,width=8.2cm}
    \hfill\mbox{ }
    \caption{\it Time evolution of the neutrino number density and the
     lepton asymmetry for the $SU(5)\times U(1)_F$ model. 
     The solid line shows the solution of the Boltzmann equation for the 
     right-handed neutrinos, while the corresponding equilibrium 
     distribution is represented by the dashed line.
     The absolute value of the lepton asymmetry $Y_L$ 
     is given by the dotted line and the hatched area shows the
     lepton asymmetry corresponding to the observed baryon asymmetry.
     \label{asyB}}
  \end{figure} 

In this case one obtains from eqs.~(\ref{cpa}) and (\ref{yuk1}),
\beq
  \ve_1\ \sim\ {3\over 16\pi}\ \e^4\;.
  \label{su5epsilon}
\eeq 
From eq.~(\ref{basym}), $\e^2 \sim 1/300$ (\ref{exp1}) and $g_* \sim 100$ 
one then obtains the baryon asymmetry,
\beq\label{est1}
Y_B \sim \k\ 10^{-8}\;.
\eeq
For $\k \sim 0.1\ldots 0.01$ this is indeed the correct order of magnitude!
The baryogenesis temperature is given by the mass of the lightest of the
heavy Majorana neutrinos,
\beq
T_B \sim M_1 \sim \e^4 M_3 \sim 10^{10}\ \mbox{GeV}\;.
\eeq
For this model, where the CP asymmetry is determined by the mass 
hierarchy of light and heavy Majorana neutrinos, baryogenesis has been 
studied in detail in \cite{bp96}. 
The generated baryon asymmetry does not depend on the flavour mixing of the 
light neutrinos, in particular the $\n_\m-\n_\t$ mixing angle. 

The solution of the full Boltzmann equations is shown in fig.~\ref{asyB} 
for the non-supersymmetric case; the supersymmetric model has been studied
in \cite{plu98}. The initial condition at a temperature $T \sim 10 M_1$ is 
chosen to be a state without heavy neutrinos. The Yukawa interactions are 
sufficient to bring the heavy neutrinos into thermal equilibrium. The 
approach of the heavy neutrino number density to the equilibrium density
and the evolution of the lepton asymmetry are analogous to GUT baryogenesis
with heavy bosons \cite{tf81}.
At temperatures $T\sim M_1$ the familiar out-of-equilibrium decays sets in,
which leads to a non-vanishing baryon asymmetry. The final asymmetry agrees 
with the estimate (\ref{est1}) for $\k \sim 0.1$. The dip in fig.~\ref{asyB} 
is due to a change of sign in the lepton asymmetry at $T \sim M_1$.

\subsubsection{$SU(3)_c\times SU(3)_L\times SU(3)_R\times U(1)_F$}

\begin{figure}[t]
    \mbox{ }\hfill
    \epsfig{file=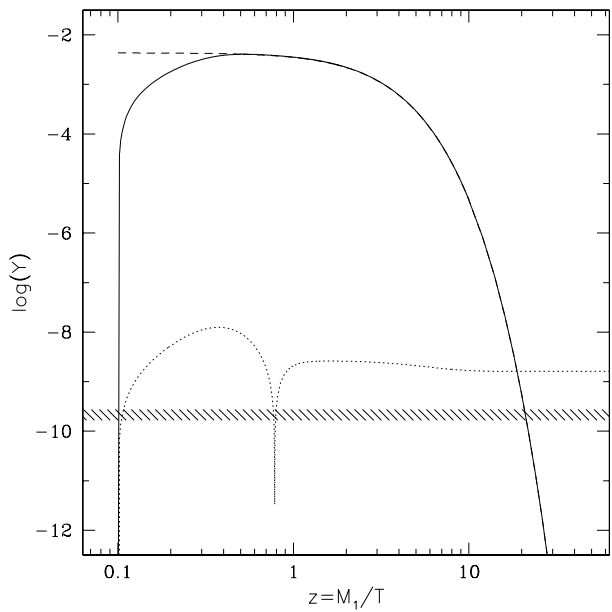,width=8.2cm}
    \hfill\mbox{ }
    \caption{\it Solution of the Boltzmann equations for the
     $SU(3)_c\times SU(3)_L\times SU(3)_R\times U(1)_F$ model. 
     \label{asyLR}}
  \end{figure} 

In this model the neutrino Yukawa couplings (\ref{yuk2}) yield the CP asymmetry
\beq
\ve_1\ \sim\ {3\over 16\pi}\ \e^5\;,
\eeq 
which correspond to the baryon asymmetry (cf.~(\ref{basym}))
\beq\label{est2}
Y_B \sim \k\ 10^{-6}\;.
\eeq
Due to the large value of $\e$ the CP asymmetry is two orders of magnitude
larger than in the $SU(5)\times U(1)_F$ model. However, 
washout processes are now also stronger. The solution of the Boltzmann 
equations is shown in fig.~\ref{asyLR}. The final asymmetry is again 
$Y_B \sim 10^{-9}$ 
which corresponds to $\k \sim 10^{-3}$. The baryogenesis temperature is
considerably larger than in the first case,
\beq
T_B \sim M_1 \sim \e^5 M_3 \sim 10^{12}\ \mbox{GeV}\;.
\eeq

The baryon asymmetry is largely determined by the parameter $\wt m_1$
defined in eq.~(\ref{ooeb}) \cite{plu97}. In the first example, one
has $\wt m_1 \sim \Bar m_\n$. In the second case one finds  
$\wt m_1 \sim m_3$. Since $\Bar m_\n$ and $m_3$ are rather similar
it is not too surprizing that the generated baryon asymmetry is about
the same in both cases.

\section{Implications for dark matter}

The experimental evidence for small neutrino masses, the see-saw
mechanism and the out-of-equilibrium condition for the decay of the heavy 
Majorana neutrinos are all consistent and suggest rather large heavy neutrino
masses and a correspondingly large baryogenesis temperature. For thermal
leptogenesis models one typically finds \cite{bp00},
\beq
T_B\ \sim\ M_1 > \ 10^7\ \mbox{GeV}\;.
\eeq

In the particularly attractive supersymmetric version of thermal leptogenesis 
one then has to consider the following two issues: the consistency of
the large baryogenesis temperature with the `gravitino constraint' and
the size of other possible contributions to the baryon asymmetry. A large 
asymmetry can in principle be generated by coherent oscillations of scalar 
fields which carry baryon and lepton number \cite{ad85}. It appears likely, 
however, that the interactions of the right-handed neutrinos are sufficiently 
strong to erase such primordial baryon and lepton asymmetries before thermal 
leptogenesis takes place \cite{bjp}. 
\begin{figure}[b]
\begin{center}
\mbox{ }\hfill
\epsfig{file=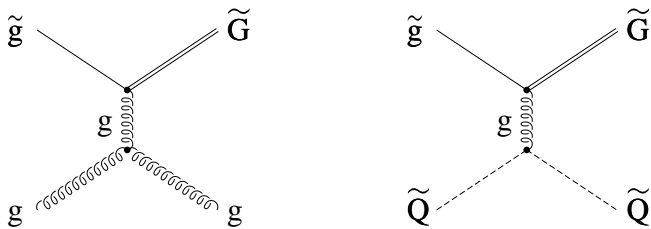,width=9cm}
\hfill\mbox{ }
\end{center}
\caption{\it Typical gravitino production processes mediated by gluon 
exchange. \label{gravprod}}
\end{figure}

The couplings of gravitinos with matter are essentially model independent. 
Their cosmological effects therefore provide very interesting information 
about possible extensions of the standard model.
It was realized long ago that standard cosmology requires gravitinos to
be either very light, $\mgr<1$\,keV \cite{pp82}, or very heavy, 
$\mgr> 10$\,TeV \cite{wei82}. These constraints are relaxed if the 
standard cosmology is extended to include an inflationary phase 
\cite{kl84,ekn84,mmy93}.
The cosmologically relevant gravitino abundance is then created in the 
reheating phase after inflation in which a reheating temperature $T_R$ is
reached. Gravitinos are dominantly produced by inelastic $\ttt$ scattering 
processes of particles from the thermal bath. The gravitino abundance is 
essentially linear in the reheating temperature $T_R$. 
It is intriguing that for temperatures $T_R \sim 10^{10}$~GeV, which are of
interest for leptogenesis, gravitinos with mass of the electroweak scale, 
$\mgr \sim 100$~GeV, can be the dominant component of cold dark matter 
\cite{bbp98}. 

The bounds on the reheating temperature depend on the thermal gravitino
production rate which is dominated by two-body processes involving gluinos 
($\gl$) (cf.~fig.~\ref{gravprod}). 
On dimensional grounds the production rate has the form
\beq
\G(T)\ \propto\ {1\over M^2}\ T^3\;,
\eeq
where $M = (8\pi G_N)^{-1/2} = 2.4\cdot 10^{18}$ GeV is the Planck mass. 
Hence, the density of thermally produced gravitinos increases strongly with 
temperature.

The production rate depends on the ratio $\mgl / \mgr$, the ratio
of gluino and gravitino masses. The ten $\ttt$ gravitino production processes 
were considered in \cite{ekn84} for $\mgl \ll \mgr$. The case $\mgl \gg \mgr$,
where the goldstino contribution dominates, was studied in \cite{mmy93}.
Four of the ten production processes are logarithmically singular due to the 
exchange of massless gluons. The complete result for the logarithmically 
singular
part of the production rate was obtained in \cite{bbp98}. The correct finite
part can be obtained by means of a hard thermal loop resummation, which was
first implemented in the case of axion production in a QED plasma \cite{by91}
and recently also for gravitino production in a QCD plasma \cite{bbb00}. 
The result for the Boltmann collision term reads
\bea\label{eq:collgrav}
C_{\gr}(T) 
&=& \factor \frac{3\zeta(3) g^2 (N^2-1)T^6}{32\pi^3 M^2}\nonumber\\
&& \hspace{.5cm}
\times\Bigg\{\left[\ln\left(\frac{T^2}{m_g^2}\right) + 0.3224\right] (N+n_f) +
0.5781 n_f \Bigg\}\;,
\eea
where
\beq
  \label{eq:gluonmass}
  m_g^2 = \frac{g^2 T^2}{6}(N + n_f)
\eeq
is the thermal gluon mass squared; $N$ is the number of colours and 
$n_f$ is the number of colour triplet and anti-triplet chiral multiplets.
The QCD coupling $g(T)\simeq 0.85$ for $T \sim 10^{10}$~GeV. For 
the supersymmetric standard model with $N_C=3$ and $n_f=6$ this implies 
$m_g > T$. Hence, the usually assumed separation of scales, 
$g^2 T \ll gT \ll T$, appears problematic and higher-order corrections
may be important.

\begin{figure}
  \begin{center}
  \epsfig{file=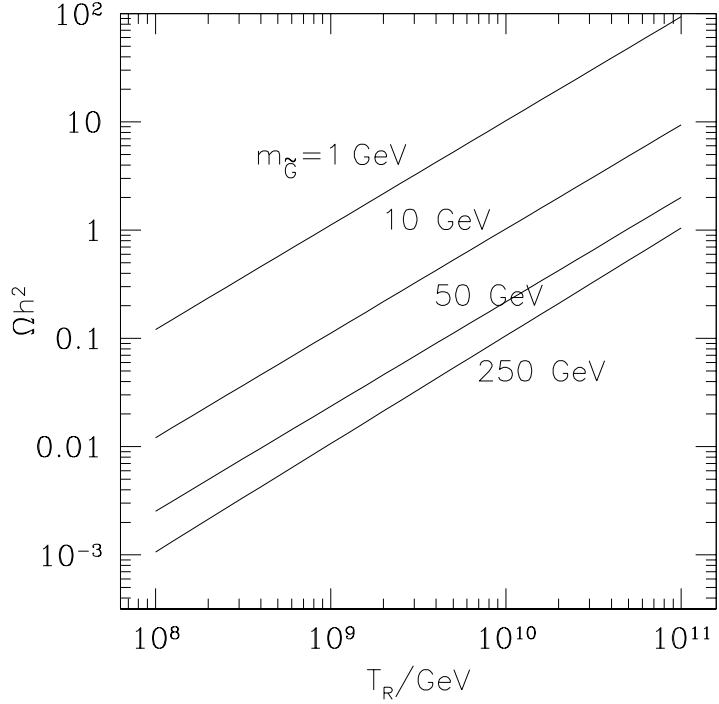,width=25em}
\caption{{\it Contribution of gravitinos to the density parameter $\Ogr h^2$ 
for different gravitino masses $\mgr$ as function of the reheating temperature
$T_R$. The gluino mass has been set to $\mgl=700$~GeV} \protect\cite{bbb00}.}
  \label{fig:omega}
  \end{center}  
\end{figure}
Using the Boltzmann equation, 
\beq
    \frac{d\ngr}{dt} + 3 H \ngr = C_{\gr}\;,
    \label{eq:Beq}
\eeq
one can calculate the gravitino abundance at temperatures $T<T_R$, 
assuming constant entropy. One finds 
\beq
  \label{eq:ygra1}
  \Ygr(T) = \frac{\ngr(T)}{\nrad(T)} \simeq 
  \frac{g_{\star S}(T)}{g_{\star S}(T_R)}
    \frac{C_{\gr}(T_R)}{H(T_R)\nrad(T_R)}\;,
\eeq
where $g_{\star S}(T)$ is the number of effectively massless degrees
of freedom \cite{rpp00}. For $T<1$~MeV, i.e. after nucleosynthesis, 
$g_{\star S}(T)=\frac{43}{11}$, whereas $g_{\star S}(T_R)=\frac{915}{4}$ 
in the supersymmetric standard model.
With $ H(T) = (g_\star(T)\pi^2/90)^{1/2} T^2/M$ 
one obtains in the case of light gravitinos 
($\mgr\ll \mgl(\mu)$, $\mu\simeq 100$~GeV)
from eqs.~(\ref{eq:ygra1}) and (\ref{eq:collgrav})
for the gravitino abundance and for the contribution to $\O h^2$, 
\beq
    \Ygr = 1.1\cdot 10^{-10}
    \left(\frac{T_R}{10^{10}\,\mbox{GeV}}\right)
    \left(\frac{100\,\mbox{GeV}}{m_{\gr}}\right)^2
    \left(\frac{\mgl(\mu)}{1\,\mbox{TeV}}\right)^2,
    \label{eq:ygra2}
\eeq
\bea
    \Ogr h^2 & = & \mgr \Ygr(T) \nrad(T) h^2 \rho_c^{-1} \nonumber \\
    & = & 0.21
    \left(\frac{T_R}{10^{10}\,\mbox{GeV}}\right)
    \left(\frac{100\,\mbox{GeV}}{\mgr}\right)
    \left(\frac{\mgl(\mu)}{1\,\mbox{TeV}}\right)^2.    
\label{eq:omgr1}
\eea  
Here we have used $\nrad(T)= \zeta(3)T^3/\pi^2$, 
and $\mgl(T)= g^2(T)/g^2(\mu) \mgl(\mu)$. 
The new result for $\Ogr h^2$ is smaller by a factor of 3 compared to the
result given in \cite{bbp98}. This is due to a partial cancellation between
the logarithmic term and the constant term in eq.~(\ref{eq:collgrav}).

It is remarkable that reheating temperatures 
$T_R = 10^8 - 10^{10}$~GeV lead to values 
$\Ogr h^2=0.01\dots 1$ in an interesting gravitino mass range. 
This is illustrated in fig.~\ref{fig:omega} for a gluino mass 
$\mgl = 700$~GeV. As an example, for $T_R\simeq 10^{10}$~GeV, 
$\mgr \simeq 80$~GeV and $h \simeq 0.65$ \cite{rpp00} one finds 
$\Ogr=0.35$, which agrees with
recent measurements of the matter density $\O_{\rm M}$ \cite{rpp00}. 

\begin{figure}
  \begin{center}
  \epsfig{file=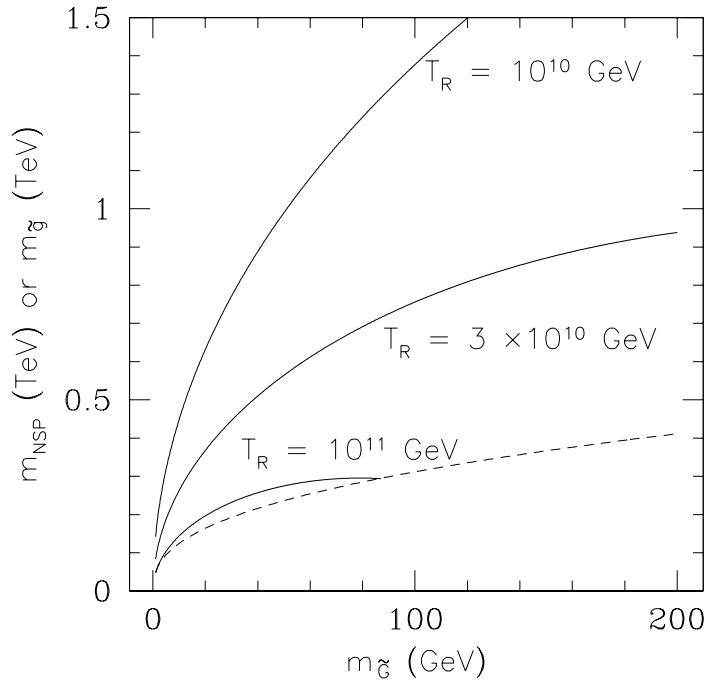,width=25em}
\caption{{\it Upper and lower bounds on the gluino mass and the NSP mass
as functions of the gravitino mass. The full lines represent the upper bound 
on the gluino mass $\mgl > m_{NSP}$ for different reheating 
temperatures from the closure limit constraint. 
The dashed line is the lower bound on $m_{NSP}$ which follows
from the NSP lifetime} \protect\cite{bbb00}.}
  \label{fig:masses}
  \end{center}  
\end{figure}

In general, to find a viable cosmological scenario one has to avoid two 
types of gravitino problems: For unstable gravitinos their decay products
must not alter the observed abundances of light elements in the universe, 
which is referred to as the big bang nucleosynthesis (BBN) constraint.
For stable gravitinos this condition has to be met by other super particles, 
in particular the next-to-lightest super particle (NSP), which decays into
gravitinos; further, their contribution to the energy density of the universe
must not exceed the closure limit, i.e. 
$\Ogr = \rho_{\gr}/\rho_c < 1$, where 
$\rho_c=3H_0^2M^2=1.05 h^2 10^{-5}$\,GeV\,cm$^{-3}$ is the critical 
energy density. 

Consider first the constraint from the closure limit. 
The condition $\Ogr = \Ygr \mgr \nrad/\rho_c \leq 1$
yields a boundary in the $\mgr$-$\mgl$ plane which is shown
in fig.~\ref{fig:masses} for three different values of the reheating
temperature  $T_R$. The allowed regions are below the three solid lines, 
respectively.

With respect to the BBN constraint, consider some
nonrelativistic particle $X$ which decays into electromagnetically and strongly
interacting relativistic particles with a lifetime $\tau_X$. Roughly
speaking, the decay changes the abundances of light elements the more
the longer the lifetime $\tau_X$ and the higher the energy density 
$m_X Y_X$ are. These constraints have been studied in detail by several groups 
\cite{egX92,km95,hkX99}. For most supergravity models they rule out the 
possibility of unstable gravitinos with $\mgr \sim 100$~GeV for 
$T_R \sim 10^{10}$~GeV, although even larger reheating temperatures are
acceptable in some cases \cite{ay00}.
\begin{figure}[tb]
\begin{center}
    \vskip .1truein
    \centerline{\epsfysize=10cm {\epsffile{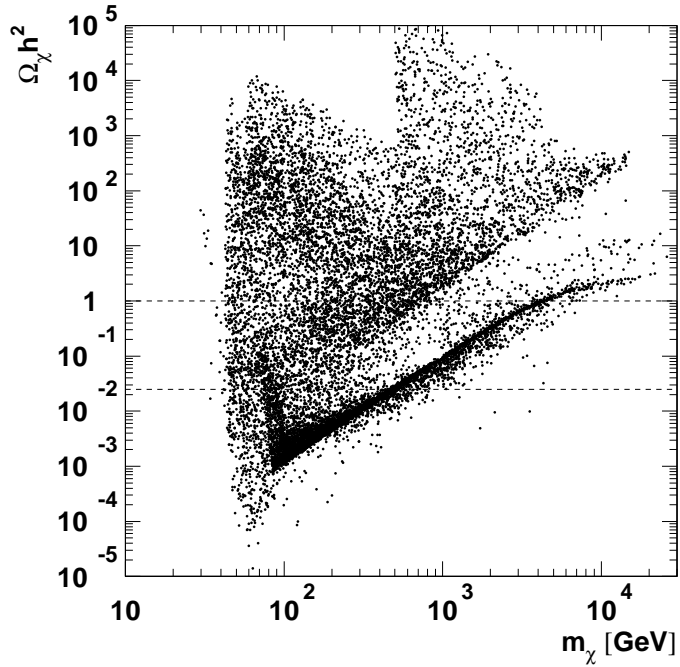}}}
    \vskip -0.1truein
\caption{{\it Neutralino relic density versus neutralino mass. The horizontal 
lines bound the region $0.025<\O_{\chi}h^2<1$} \protect\cite{eg97}.}
\label{gondo}
\end{center}
\vspace{-0.5cm}
\end{figure}

For stable gravitinos the NSP plays the role 
of the particle X. The lifetime of a fermion decaying into its scalar partner 
and a gravitino is 
\beq
\tau_{\rm NSP} = 48\pi \frac{\mgr^2 M^2}{m_{\rm NSP}^5}.
\eeq
For a sufficiently short lifetime, 
$\tau_{\rm NSP}<2\cdot 10^6\,\mbox{s}$, it is sufficient to require the 
energy density which becomes free in NSP decays to be smaller than 
$m_X Y_X < 4\cdot 10^{-10}\,\mbox{GeV}$, 
which corresponds to $\O_X h^2 < 0.008$. 
This constraint can be satisfied since the NSP relic density is rather
model dependent. For neutralinos in the MSSM 
the energy fraction $\O_{\chi}$
can vary over eight orders of magnitude (cf.~fig.~(\ref{gondo})).
The lifetime constraint $\tau_{\rm NSP}<2\cdot 10^6\,\mbox{s}$  yields
a lower bound on super particle masses which is represented by the dashed
line in the $\mgr$-$m_{\rm NSP/\gl}$ plane in  fig.~\ref{fig:masses}. 
 
In order to decide whether the second part of the BBN constraint, 
$\O_{\rm NSP} h^2 < 0.008$,  is satisfied, one has  to 
specify which particle is the NSP. The case of a higgsino-like neutralino  
as NSP  has been studied in \cite{bbp98}. A detailed discussion for a 
scalar $\tau$-lepton as NSP has been given in \cite{ggr99},\cite{ahs00}.

\section{Outlook}

Recent developments in theoretical and experimental particle physics support
the idea of leptogenesis according to which the cosmological matter density 
has been created in decays of heavy Majorana neutrinos. On the theoretical 
side, detailed studies of the electroweak phase transition and sphaleron 
processes have shown that the matter-antimatter asymmetry is related to 
neutrino properties. On the experimental side, the solar and atmospheric 
neutrino deficits have been observed, which can be interpreted as a result of
oscillations between three species of light Majorana neutrinos. 

It is very remarkable that these hints on the nature of lepton number
violation fit very well together with the leptogenesis mechanism.  For
hierarchical neutrino masses, with $B-L$ broken at the unification
scale $\Lambda_{\mbox{\scriptsize GUT}}\sim 10^{16}\;$GeV, the
observed baryon asymmetry $n_B/s \sim 10^{-10}$ is naturally explained
by the decay of heavy Majorana neutrinos. The corresponding
baryogenesis temperature is $T_B \sim 10^{10}$ GeV.

The consistency of this picture has implications in particle physics and
cosmology. In unified theories the pattern of neutrino masses and mixings
is related to lepton flavour and quark flavour changing processes. 
In supersymmetric theories the mass spectrum of superparticles is constrained
by the cosmological bound on the number density of gravitinos which may
be the dominant component of dark matter. Further, the realization of the
rather large baryogenesis temperature in models of inflation should have
observable consequences for the anisotropy of the cosmic microwave background.

\vspace*{0.5truecm}

\begin{flushleft}
{\it Acknowledgements}
\end{flushleft}
The content of these lectures is based on work in collaboration with M.~Bolz,
A.~Brandenburg, S.~Fredenhagen, M.~Pl\"umacher and T.~Yanagida whom I thank 
for a fruitful collaboration. I am also grateful to the organizers for
arranging an enjoyable and stimulating meeting.


\newpage

\begin{thebibliography}{}  

\bibitem{boo}
P.~de~Bernardis et al., Nature {\bf 404} (2000) 955

\bibitem{rpp00}
Review of Particle Physics, Eur. Phys. J. {\bf C15} (2000) 1

\bibitem{RF}
R.~Fleischer, these proceedings

\bibitem{sa67}
A.~D.~Sakharov, JETP Lett.~{\bf 5} (1967) 24

\bibitem{tho76}
G.~'t~Hooft, \prl{37}{76}{8}; \pr{14}{76}{3422}

\bibitem{yo78}
M.~Yoshimura, \prl{41}{78}{281}; {\it ibid.} {\bf 42} (1979) 746 (E);\\
S.~Dimopoulos, L.~Susskind, \pr{18}{78}{4500};\\
D.~Toussaint, S.~B.~Treiman, F.~Wilczek, A.~Zee, \pr{19}{79}{1036};\\
S.~Weinberg, \prl{42}{79}{850}

\bibitem{fy86} 
M.~Fukugita, T.~Yanagida, \pl{174}{86}{45}

\bibitem{ad85}
I.~Affleck, M.~Dine, \np{249}{85}{361}

\bibitem{CW}
C.~Wetterich, these proceedings

\bibitem{AL}
A.~Linde, these proceedings

\bibitem{rt99}
For a review and references, see\\
A.~Riotto, M.~Trodden, Ann.~Rev.~Nucl.~Part.~Sci. {\bf 49} (1999) 35

\bibitem{krs85}
V.~A.~Kuzmin, V.~A.~Rubakov, M.~E.~Shaposhnikov, \pl{155}{85}{36}

\bibitem{boe98}
D.~B\"odeker, \pl{426}{98}{351}

\bibitem{moo00}
G.~D.~Moore, {\it Do We Understand the Sphaleron Rate?}, {\tt hep-ph/0009161}

\bibitem{ht90}
J.~A.~Harvey, M.~S.~Turner, \pr{42}{90}{3344}

\bibitem{bp00}
For a review and references, see\\
W.~Buchm\"uller, M.~Pl\"umacher, {\it Neutrino Masses and the Baryon 
Asymmetry}, {\tt hep-ph/0007176} 

\bibitem{kw80}
E.~W.~Kolb, S.~Wolfram, \np{172}{80}{224}; \np{195}{82}{542}(E)

\bibitem{lut92}
M.~A.~Luty, \pr{45}{92}{455}

\bibitem{plu97}
M.~Pl\"umacher, Z.~Phys.~{\bf C\ 74} (1997) 549;

\bibitem{jmy98}
I.~Joichi, S.~Matsumoto, M.~Yoshimura, \pr{58}{98}{43507}

\bibitem{bf00}
W.~Buchm\"uller, S.~Fredenhagen, Phys. Lett. {\bf B 483} (2000) 217

\bibitem{lan}
L.~D.~Landau, E.~M.~Lifshitz, {\it Statistical Physics}, Addison-Wesley (1959)

\bibitem{mz92}
R.~N.~Mohapatra, X.~Zhang, \pr{45}{92}{2699}

\bibitem{ks88}
S.~Yu.~Khlebnikov, M.~E.~Shaposhnikov, \np{308}{88}{885}

\bibitem{ks96}
S.~Yu.~Khlebnikov, M.~E.~Shaposhnikov, \pl{387}{96}{817}

\bibitem{ls00}
M.~Laine, M.~E.~Shaposhnikov, Phys.~Rev. {\bf D 61} (2000) 117302

\bibitem{cko93}
J.~M.~Cline, K.~Kainulainen, K.~A.~Olive, \prl{71}{93}{2372}; \pr{49}{94}{6394}

\bibitem{fy90}
M.~Fukugita, T.~Yanagida, \pr{42}{90}{1285}

\bibitem{seesaw} 
T.~Yanagida, in {\it{Workshop on unified Theories}}, KEK report 
79-18 (1979) p.~95;\\
M.~Gell-Mann, P.~Ramond, R.~Slansky, in {\it{Supergravity}} (North Holland, 
Amsterdam, 1979) eds. P.~van Nieuwenhuizen, D.~Freedman, p.~315

\bibitem{EA}
E.~Akhmedov, these proceedings

\bibitem{fn79}
C.~D.~Froggatt, H.~B.~Nielsen, \np{147}{79}{277}

\bibitem{atm98}
Super-Kamiokande Collaboration, Y.~Fukuda et al., \prl{81}{98}{1562}

\bibitem{ys99}
T.~Yanagida, J.~Sato, Nucl.~Phys.~{\bf B} Proc.~Suppl. {\bf 77} (1999) 293

\bibitem{ram99}
P.~Ramond, Nucl.~Phys.~{\bf B} Proc.~Suppl. {\bf 77} (1999) 3

\bibitem{bw87}
J.~Bijnens, C.~Wetterich, \np{292}{87}{443}

\bibitem{lr99}
S.~Lola, G.~G.~Ross, \np{553}{99}{81}

\bibitem{by99}
W.~Buchm\"uller, T.~Yanagida, \pl{445}{99}{399}

\bibitem{ilr98}
N.~Irges, S.~Lavignac, P.~Ramond, \pr{58}{98}{035003}

\bibitem{vis98}
F.~Vissani, JHEP11 (1998) 025

\bibitem{hl97}
N.~Hata, P.~Langacker, \pr{56}{97}{6107}

\bibitem{msw86}
S.~P.~Mikheyev, A.~Y.~Smirnov, \nc{9C}{86}{17};\\
L.~Wolfenstein, \pr{17}{78}{2369}

\bibitem{kt90}
E.~W.~Kolb, M.~S.~Turner, {\it The Early Universe}, Addison-Wesley,
New York, 1990

\bibitem{fgx91}
W.~Fischler, G.~F.~Giudice, R.~G.~Leigh, S.~Paban, \pl{258}{91}{45}

\bibitem{by93}
W.~Buchm\"uller, T.~Yanagida, \pl{302}{93}{240}

\bibitem{fps95}
M.~Flanz, E.~A.~Paschos, U.~Sarkar, \pl{345}{95}{248}; \pl{384}{96}{487} (E)

\bibitem{crv96}
L.~Covi, E.~Roulet, F.~Vissani, \pl{384}{96}{169}

\bibitem{bp98}
W.~Buchm\"uller, M.~Pl\"umacher, \pl{431}{98}{354}

\bibitem{pil99}
For a discussion and references, see\\
A.~Pilaftsis, Int.~J.~Mod.~Phys. {\bf A14} (1999) 1811

\bibitem{GL}
G.~Lazarides, these proceedings

\bibitem{bp96} 
W.~Buchm\"uller, M.~Pl\"umacher, \pl{389}{96}{73}

\bibitem{plu98}
M.~Pl\"umacher, \np{530}{98}{207}

\bibitem{tf81}
M.~S.~Turner, J.~N.~Fry, \pr{24}{81}{3341}

\bibitem{bjp}
W.~Buchm\"uller, A.~Jakov\'ac, M.~Pl\"umacher, in preparation

\bibitem{pp82}
  H.~Pagels, J.~R.~Primack, \prl{48}{82}{223}

\bibitem{wei82}
  S.~Weinberg, \prl{48}{82}{1303}

\bibitem{kl84}
  M.~D.~Khlopov, A.~D.~Linde, \pl{138}{84}{265}

\bibitem{ekn84}
  J.~Ellis, J.~E.~Kim, D.~V.~Nanopoulos, \pl{145}{84}{181}

\bibitem{mmy93}
T.~Moroi, H.~Murayama, M.~Yamaguchi, \pl{303}{93}{289}

\bibitem{bbp98}
  M.~Bolz, W.~Buchm\"uller, M.~Pl\"umacher, \pl{443}{98}{209}

\bibitem{by91}
E.~Braaten, T.~C.~Yuan, \prl{66}{91}{2183}

\bibitem{bbb00}
M.~Bolz, A.~Brandenburg, W.~Buchm\"uller, {\tt hep-ph/0012052}

\bibitem{egX92}
J.~Ellis, G.~B.~Gelmini, J.~L.~Lopez, D.~V.~Nanopoulos, S.~Sarkar,
  \np{373}{92}{399}

\bibitem{km95}
M.~Kawasaki, T.~Moroi, \ptp{93}{95}{879}

\bibitem{hkX99}
E.~Holtmann, M.~Kawasaki, K.~Kohri, T.~Moroi, \pr{60}{99}{023506}

\bibitem{ay00}
T.~Asaka, T.~Yanagida, Phys.~Lett. {\bf B494} (2000) 297

\bibitem{ggr99}
T.~Gerghetta, G.~F.~Giudice, A.~Riotto, \pl{446}{99}{28}

\bibitem{ahs00}
T.~Asaka, K.~Hamaguchi, K.~Suzuki, Phys.~Lett. {\bf B490} (2000) 136

\bibitem{kk00}
R.~Kallosh, L.~Kovman, A.~Linde, A.~van Proeyen, 
Phys.~Rev. {\bf D61} (2000) 103503

\bibitem{grt99}
G.~F.~Giudice, A.~Riotto, I.~Tkachev, JHEP 9911 (1999) 036

\bibitem{eg97}
  J.~Edsj\"o, P.~Gondolo, \pr{56}{97}{1879}


\end{thebibliography}
\end{document}